\documentclass[onecolumn,floats,floatfix,amssymb,nofootinbib,prd,superscriptaddress]{revtex4-1}

\usepackage{graphicx}
\usepackage{amsmath,amssymb}
\usepackage{amsfonts}
\usepackage{xspace} 
\usepackage[usenames]{color}
\usepackage{dcolumn}
\usepackage{bm}
\usepackage{mathrsfs}
\usepackage{multirow}
\usepackage[colorlinks=true]{hyperref}
\usepackage[all]{hypcap} 
\usepackage[utf8]{inputenc} 
\usepackage{lipsum}
\usepackage{etoolbox}
\usepackage{tikz}
\usepackage{url}
\usepackage{subfigure}
\usepackage{adjustbox}
\usepackage{caption}
\usepackage{slashed}
\usepackage{multirow}
\usepackage{caption}
\usepackage{array}
\usepackage{booktabs}
\usepackage{siunitx}

\usepackage{lineno}

\def\be{\begin{equation}}
\def\ee{\end{equation}}
\def\bea{\begin{eqnarray}}
\def\eea{\end{eqnarray}}

\newcommand{\beq}{\begin{eqnarray}}
\newcommand{\eeq}{\end{eqnarray}}
\usepackage{calrsfs}
\DeclareMathAlphabet{\pazocal}{OMS}{zplm}{m}{n}

\begin{document}

\title{Electrically charged black hole solutions in semiclassical gravity\\ and dynamics of linear perturbations}

\author{Adri\'an del R\'{\i}o}\email{adrian.rio@uv.es}
\affiliation{Departament de F\'isica Te\'orica and IFIC, Universitat de Valencia - CSIC  \\ Dr Moliner 50, 46100, Burjassot (Valencia), Spain}
\author{Evelyn-Andreea Ester}\email{evelyn-andreea-ester@nbi.ku.dk}
\affiliation{Niels Bohr International Academy, Niels Bohr Institute, Blegdamsvej 17, 2100 Copenhagen, Denmark} 
\begin{abstract}

We explore quantum corrections of electrically charged black holes subject to vacuum polarization effects of  fermion fields in QED.  Solving this problem exactly is challenging so we restrict to  perturbative corrections that one can obtain using the heat kernel expansion in the one-loop  effective action for electrons. Starting from the corrections originally computed by Drummond and Hathrell, we solve the full semiclassical Einstein-Maxwell  system of coupled equations to leading order in Planck's constant, and  find a new electrically charged, static black hole solution. To probe these quantum corrections, we  study  electromagnetic and gravitational (axial) perturbations on this background, and derive the coupled system of Regge-Wheeler master equations  that govern the propagation of these waves. In the classical limit our results agree with previous findings in the literature. We finally compare these results with those that one can obtain by working out the Euler-Heisenberg effective action.  We find again a new electrically charged static black hole spacetime, and derive the coupled system of Regge-Wheeler equations governing the propagation of axial electromagnetic and gravitational perturbations. Results are qualitatively similar in both cases. We  briefly discuss some  challenges found in the  numerical computation of  the quasinormal mode frequency spectra when quantum corrections are included. 

{\let\clearpage\relax  \let\sectionname\relax \tableofcontents}
\end{abstract}
\maketitle

\section{Introduction}
\label{sec:Introduction}

General relativity is one of the pillars of modern physics, yet it is not a complete description of the gravitational interaction, as it fails to resolve black hole singularities or to describe quantum aspects. To obtain a foundational description of gravity, the Einstein field equations will ultimately need modifications. The study of the semiclassical Einstein equations, which incorporates the vacuum energies and stresses of quantum fields \cite{birrell_davies_1982,Wald:1995yp,parker_toms_2009}, can be helpful as a first approach to explore these modifications.

From a theoretical standpoint, the coalescence of two black holes (BHs)  is not only a fascinating process in classical general relativity \cite{Barack:2018yly}, but can also be a rich laboratory as to how quantum field theories work, which displays their consistency issues in new and edifying situations. In order to test new fundamental physics with gravitational waves from coalescences of compact objects, it is important to know what results to expect and how to model them to the required precision. Ideally, one wants to obtain the maximal amount of information from the data, instead of looking for very specific predictions, in order not to miss unexpected new physics. Effective field theories \cite{Burgess:2003jk} are at an advantage, because their range of validity is broad, so they allow one to combine constraints coming from the strong gravity regime with bounds from e.g. the weak field regime, cosmology, astrophysics, laboratory tests, etc. One drawback, however, is that this approach requires working with a free set of parameters along calculations. In the present work,  given the complications of some of the equations, we focus for simplicity on the predictions of quantum electrodynamics (QED) in curved spacetimes, as it is the oldest, simplest and most successful quantum field theory. 

Once this framework is fixed, we can focus on computing the corrections to BH solutions and their dynamics. In a BH merger, an especially interesting stage concerns the late time dynamics, described by a ``ringdown'' phase, during which the distorted remnant sheds its nontrivial multipolar structure through gravitational waves (or other radiation), and relaxes to the final stationary solution. During this stage, the dynamics is well described by a set of quasinormal modes (QNMs) of the final stationary solution, characterized by complex characteristic  frequencies~\cite{Kokkotas:1999bd,Nollert_1999,Berti:2009kk,Konoplya:2011qq} (but nonlinearities, initial transients and backscattering may also play a role~\cite{Baibhav_2023,Zhu_2024,Cheung:2022rbm,Mitman:2022qdl}). So far, gravitational-wave observations involving BHs and their dynamics are well described by classical general relativity with great accuracy (i.e., by the Kerr solution), and any quantum correction is {\it a priori} too  suppressed to be observed with astrophysical measurements. Nevertheless, it is important to understand what our fundamental theories entail, and how to build on their ideas, even if, in practice, their predictions may be elusive for current experiments.  Similarly, great effort was made in the mid-$20${th} century to understand QED  to all orders in perturbation theory, despite the fact that only the first few orders could be corroborated in accelerators at that time. Moreover, sometimes  apparently slight perturbations of the background can lead to drastic changes. Known examples concern the superradiant instability of Kerr geometries against massive fields~\cite{Brito:2015oca}; the spectral instability of BHs under small changes of the effective potential governing massless fluctuations~\cite{Nollert:1996rf, PhysRevD.101.104009, Barausse:2014tra,Jaramillo:2020tuu,Cheung:2021bol}; or the disappearance of the BH horizon in  nonperturbative calculations within semiclassical gravity~\cite{Beltran-Palau:2022nec}.  Therefore, the topic deserves further study.

In this work we attempt to explore the vacuum polarization effects of fermion fields on BH solutions in general relativity. The implications of vacuum polarization can be studied from the one-loop effective action.  Arguably, the most renowned example is the Euler-Heisenberg Lagrangian in QED. Roughly speaking, this is the result of ``integrating out'' the dynamics of the electrons in the QED action, producing as a consequence a new Lagrangian that  describes the effective dynamics of the electromagnetic field $F_{ab}$. This yields a nonlinear theory whose leading-order corrections, for {\it sufficiently weak fields}, are given by \cite{DUNNE_2005, parker_toms_2009}
\beq
\label{eq:Euler-HeisenbergLagrangian}
{\cal L}_\text{eff}[A] &=&
-\frac{1}{4}F_{ab}F^{ab}-\frac{\hbar e^4}{45(4\pi)^2m_e^4}\left[\frac{5}{4}(F_{ab}F^{ab})^2-\frac{7}{2}F_{ab}F_{cd}F^{ac}F^{bd}\right]\,, 
\eeq
where $m_e$ is the electron mass, $e$ is the electron charge, and $A$ is the electromagnetic potential (defined by $F=dA$). The free Maxwell action, written in the first line above, gets corrected by an extra, highly nonlinear contribution, which captures the backreaction of quantum vacuum fluctuations of the fermion field, and which modifies the classical dynamics of the electromagnetic field.

Effective actions  allow one to explore a theory like QED from a different perspective. In particular, one-loop effective actions are able to capture nonperturbative phenomena. To give an example, the full Euler-Heisenberg effective Lagrangian  predicts the Schwinger effect in the strong field limit  (i.e. the excitation of electron-positron pairs out of the quantum vacuum by a strong electric field), which is otherwise not derivable from any order in perturbative QED \cite{DUNNE_2005}. 

We wish to analyze the similar problem in general relativity. In close analogy to QED, one first attempt is to study the (backreaction) effects that the quantum vacuum  of a fermion field can produce on a  spacetime metric. Renormalizability arguments \cite{birrell_davies_1982} yield an effective Einstein-Hilbert action that, to leading order, reads
\bea
{\cal L}_\text{eff}[g]= \frac{R}{16\pi}+\hbar\left(\alpha_1 R^2+\alpha_2 R_{ab}R^{ab}\right)+\mathcal O(\hbar^2), \label{firsttry}
\eea
for some real  numbers $\alpha_1$, $\alpha_2$. The usual Einstein-Hilbert Lagrangian $\frac{R}{16\pi}$ gets corrected by higher-order derivative contributions, which account for the vacuum polarization effects of the electrons. The presence of these terms, regardless of how small the prefactors $\alpha_1$, $\alpha_2$ might be, makes the new theory considerably different from the original: additional families of solutions arise, some of which are ``runaways'', i.e. differ drastically from the original theory. 

Higher-order derivative contributions in the field equations typically arise as a result of truncating a perturbative expansion of a nonlocal theory, which does not suffer from these problems. If one expects that quantum corrections will not dramatically change the behavior of the classical system, then perturbative constraints must be applied on the action to disregard solutions that do not exist in the limit of a zero-expansion parameter, $\hbar\sim 0$  \cite{Simon:1990ic,Simon:1990jn}. This requirement is what ensures that the  series expansion in the action can be considered as a legitimate perturbative expansion of some complicated, nonlocal functional of the metric, which is expected from the UV completion of the theory. The perturbative constraints consist of imposing the leading-order equation of motion, which in this case is $R_{ab}=0$. This has the effect that the new (constrained) field equations ignore the $\propto R^2, R_{ab}R^{ab}$ piece entirely in our problem\footnote{Alternatively, the freedom to perform field reparametrizations allows one to get rid of all these higher-order terms in (\ref{firsttry}) for vacuum gravity, see e.g. \cite{Burgess:2003jk}} ---although this can dramatically change if matter fields are included. Therefore, the above action (\ref{firsttry}) becomes trivial with these constraints, i.e. to leading order the spacetime metric does not ``sense'' the electron's vacuum fluctuations.

To get a nontrivial problem it is necessary to consider a case that, in the classical limit, does not satisfy the vacuum Einstein equations. A natural possibility is to consider both electromagnetic and gravitational backgrounds. In this case, the effective action contains many more  terms, because it depends on two fields: $A_{a}$ and $g_{ab}$, and mixed combinations are allowed. An explicit expression of the one-loop effective action was derived by Drummond and Hathrell in Ref.~\cite{Drummond:1979pp}, where the leading-order corrections for weak fields, of order $\mathcal O\left(\hbar \frac{e^2}{m_e^2}\right)$,  were obtained using different methods. As we will see in more detail in the next sections, the presence of these corrections can produce interesting results.

In this paper, we will look for static, spherically symmetric exact solutions to the {\it full} Einstein-Maxwell system of field equations at one-loop order in the quantum corrections. We will derive first the new BH solution taking into account the Drummond-Hathrell corrections only. Then, we will do a similar analysis to obtain the BH solution with Euler-Heisenberg corrections, which dominate in the regime of high electric fields. The resulting expressions can be understood as  the extension of the Reissner-Nordstr\"{o}m  BHs of mass $M$ and charge $Q$ to the semiclassical regime, subject under the influence of the vacuum polarization of electrons.  For $Q=0$ quantum corrections at first order in Planck's constant vanish and we recover the usual Schwarzschild neutral spacetime, as expected from the discussion above.

In order to probe the properties of these two  BH backgrounds,  we will further analyze the propagation of electromagnetic and gravitational linear perturbations around, and examine the way the underlying quantum corrections can leave imprints on gravitational-/electromagnetic-wave observations. 
This is technically a rather involved calculation because the field equations couple both types of perturbations. As a first approach, we first consider the $Q=0$ case for the Drummond-Hathrell BH. In this case, not only do the electromagnetic and gravitational perturbations decouple, but the resulting wave equation  for  electromagnetic perturbations  still receives quantum corrections, so it is an interesting problem on its own (for gravitational perturbations we just recover the classical Regge-Wheeler/Zerilli equations for axial/polar  perturbations \cite{Berti:2009kk}). We  derive the effective Regge-Wheeler/Zerilli equations for the propagation of both axial and polar electromagnetic perturbations. Then, we  compute the new QNM frequencies, and we  find that quantum corrections break the well-known classical isospectrality, i.e. the vacuum polarization acts differently on axial and polar waves. In this sense, we will point out and correct some previous statements made in the literature. 

Once this first approach is well understood, we move on to address the full problem with the help of the package $\texttt{xAct}$ for {\it Mathematica}. We will provide  the relevant coupled differential equations governing the evolution of gravitational/electromagnetic axial perturbations propagating on both Drummond-Hathrell and Euler-Heisenberg charged BH backgrounds. {We will end by discussing several difficulties that we found when trying to compute the QNM frequency spectra.}

 The outline of this article is as follows. In Sec. \ref{sec:prel} we set up the necessary theoretical formalism underlying one-loop effective actions. In Sec. \ref{DHsolution}, we obtain a novel BH solution of mass $M$ and charge $Q$ of the full semiclassical Einstein-Maxwell system of equations, derived from the Drummond-Hathrell corrections and to leading-order in Planck  constant.  To probe this solution,  in Sec. \ref{sec:partialproblem} we first focus on the $Q=0$ case, and calculate  the wave equation for electromagnetic perturbations around this background solution, as well as the first few characteristic QNM frequencies.   After this, in Sec. \ref{sec:DH} we address the full problem of studying  electromagnetic and gravitational linear perturbations on this new electrically charged BH background. We restrict only to axial perturbations, for simplicity. Then, in Sec. \ref{sec:EH} we obtain again a novel BH solution of the semiclassical Einstein-Maxwell equations, but now restricting to the Euler-Heisenberg corrections of the one-loop effective action. We study axial, coupled gravitational and electromagnetic perturbations on this new background solution in Sec. \ref{sec:EH}, and we finalize the article by discussing some final remarks in Sec. \ref{sec:Conclusions}.

 Throughout this article we work with the system of units $G=c=1$. To emphasize quantum effects we will keep Planck's constant $\hbar\neq 1$ explicit. Our metric signature will be $(-,+,+,+)$, $\nabla_a$ will represent the Levi-Civita connection of the metric $g_{ab}$.{ The other sign conventions conform with Ref.~\cite{Wald:1984rg}. We use the {\it Mathematica} package $\texttt{xAct}$ \cite{xAct}, specifically the $\texttt{xTensor}$ package \cite{xTensor} and the associated  $\texttt{xTras}$ additions \cite{Nutma:2013zea}.

\section{The effective action}\label{sec:prel}

Let us consider  a Dirac field $\Psi$, physically representing electrons and positrons of mass $m_e$, interacting with electromagnetic $F_{ab}$ and gravitational  $g_{ab}$ fields. This  theory is described by the action
\begin{widetext}
\bea \label{classicalA}
S[\Psi,g, A]=\int d^4x \sqrt{-g} \left[\frac{R}{16\pi}-\frac{1}{4}F^{ab}F_{ab}+\bar\Psi (i\hbar\gamma^a D_a-m_e \,\mathbb I) \Psi \right]\, ,
\eea
\end{widetext}
where $\gamma^a$ are the usual Dirac matrices satisfying the Clifford algebra $\{\gamma^a, \gamma^b\}=2g^{ab}$, and $R$ is the Ricci scalar curvature. The Dirac field $\Psi$ couples to both backgrounds through the covariant derivative $D_a=\nabla_a +i \frac{e}{\hbar} A_a$, where $e$ is the electron charge, $A_a$ is the electromagnetic potential defined by $F=dA$, and $\nabla_a$ is the Levi-Civita connection of the metric $g_{ab}$. 

Classical electromagnetic and gravitational fields can excite or modify quantum vacuum fluctuations of the fermion field $\Psi$, and the latter  can backreact on the  fields $A_a$,  $g_{ab}$ and lead to some quantum corrections. These quantum modifications can be studied by constructing an effective action $\Gamma[\langle \Psi\rangle, g,A]$ that only depends on the variables $A_a$, $g_{ab}$ (as well as a choice of fermionic vacuum state $\langle \Psi\rangle$),   such that extremizing it with respect to them yields the semiclassical field equations:
\bea
G_{ab}=8\pi ( T^M_{ab}+\langle T_{ab} \rangle)\, , \quad \nabla_a F^{ab}=\langle J^b \rangle\, . \label{fieldeqns0}
\eea
In these equations $\langle T_{ab} \rangle$, $\langle J^b \rangle$ are the vacuum expectation values of the stress-energy tensor and electric current produced by the fermion field $\Psi$, and   $T^M_{ab}$ represents the usual source-free Maxwell stress-energy tensor:
\bea
T^M_{ab}=F_{ca}F^{c}_{\, \, b}- \frac{1}{4}g_{ab}F_{cd}F^{cd} \label{classicalST}\, .
\eea

Details on the derivation of this effective action can be consulted in standard textbooks (see e.g. chapter 6 in \cite{parker_toms_2009}). A standard strategy consists of obtaining a (formal) Feynman path integral representation  for the effective action, where the fermionic degrees of freedom are integrated out from the classical action (\ref{classicalA}). However, in general the resulting  expression  only produces an implicit equation for $\Gamma[\langle \Psi\rangle, g,A]$, which appears on both sides of the  equality. The usual method of computation is to resort then to a perturbative expansion of $\Gamma[\langle \Psi\rangle, g,A]$ in powers of the Planck's constant $\hbar$, and to solve this path integral representation  iteratively.  This perturbative expansion is called the loop expansion because each term in the expansion admits an interpretation in terms of Feynman diagrams.

At one-loop order the effective action formally reads 
\bea \label{1loopEA}
\Gamma^{(1)}[\langle \Psi\rangle, g,A]&=& S[g,A]- \frac{\hbar}{2} \int d^4 x \sqrt{-g} \int_0^{\infty} \frac{d s}{s} \operatorname{Tr} K(s; x, x) e^{- s \frac{m_e^2}{\hbar^2}}\, ,
\eea 
where 
\bea 
S[g,A]=\int d^4x \sqrt{-g} \left[\frac{R}{16\pi}-\frac{1}{4}F^{ab}F_{ab}\right] \label{freeA}\, .
\eea
 is the free action for gravity and electrodynamics, and $K(s;x,x')$ is the heat kernel of the Dirac field. The heat kernel is a bidistributional solution of $\partial_s K(s;x,x')=-\Delta_x\, K(s;x,x')$ with initial data  $K(0;x,x')=\delta^{(4)}(x-x') \mathbb I$,  where $\Delta_x=(-i\gamma^a D_a)(i\gamma^b D_b)$ is a second-order differential operator.  The specification of a vacuum state for the Dirac field provides the necessary boundary conditions to solve this equation.

In general, the heat kernel equation cannot be solved in full, closed  form. However, it is still possible to extract partial information from $K(s;x,x)$, independent of the choice of boundary conditions, which can be used to obtain a first approximate expression for (\ref{1loopEA}). The exponential suppression in equation (\ref{1loopEA}) reveals that the dominant contribution to the integral comes from the $s\lesssim \lambda_e^2$ regime, where  $\lambda_e=\frac{\hbar}{m_e}$ is the Compton wavelength of the electron. This is known as the ultraviolet (UV) regime of the Dirac theory. Interestingly, in the UV limit $s\to 0$, the  heat kernel admits an asymptotic  expansion of the form\footnote{For manifolds without boundaries all odd orders vanish, $E_{2k+1}(x)=0$ \cite{Vassilevich2003}. } 
\bea
K(s ; x, x) \underset{s\to  0}{\sim} (4 \pi  s)^{-2} \sum_{k=0}^{\infty}s^{k} E_{2k}(x) \label{HKexpansion}\, ,
\eea
where $E_{k}(x)$ are called the heat kernel coefficients  \cite{Vassilevich2003}. Each $E_k(x)$ in the perturbation series is a linear combination of spinor-valued matrices that are constructed out of contractions of the Riemann tensor  $R_{abcd}$ and the electromagnetic field strength $F_{ab}$. Furthermore, they are independent of the choice of vacuum state, or boundary conditions  for the heat kernel equation. For instance, for a differential operator of the form $\Delta_x=-\mathbb I\, g^{ab}\nabla_a \nabla_b -Q(x)$, the first few orders are \cite{parker_toms_2009, Vassilevich2003}:
\bea
E_0&=&\mathbb I \, ,\\
E_2&=&\frac{R}{6} +Q\, ,\\
E_4&=&\left(\frac{1}{30} \square R+\frac{1}{72} R^{2}-\frac{1}{180} R^{a b} R_{a b}+\frac{1}{180} R^{a b c d} R_{a b c d}\right) \mathbb I +\frac{1}{12} W^{a b} W_{a b}+\frac{1}{2} Q^{2}+\frac{1}{6} R Q+\frac{1}{6} \square Q\, ,\\
\dots\nonumber
\eea
where $W_{ab}=[\nabla_a, \nabla_b]$. In our case, it is not difficult to show that $Q=\frac{R}{4}\mathbb I+i \frac{e}{2\hbar}F_{ab}\gamma^a \gamma^b$ and $W_{ab}=\frac{1}{4}R_{abcd}\gamma^{c}\gamma^d+i \frac{e}{\hbar}F_{ab}\mathbb I$.

As a general rule, the $k$th order in the series expansion (\ref{HKexpansion}) counts  the number of  derivatives of the spacetime metric and electromagnetic potential, counting the latter as one derivative ($g_{ab}$ is said to be of zero adiabatic order, while the connections $A_a$ and $\nabla_a$ are regarded of first adiabatic order). For instance, $E_0$ has zero  derivatives and is said to be of zero adiabatic order, $E_2$ has two  derivatives and is of second adiabatic order, etc. In this sense, higher-order heat kernel coefficients in the asymptotic expansion  measure higher deviations from a ``flat'' background, where $R_{abcd}=0$ and $F_{ab}=0$. 

The first three (even) orders  in this asymptotic expansion (\ref{HKexpansion}) lead to formally UV divergent integrals in (\ref{1loopEA}), and the one-loop effective action needs to be renormalized order by order to get a finite expression. This is achieved by reabsorbing the UV divergences in  coupling constants of the classical action (\ref{freeA}) and adding suitable counterterms. The result of this process yields a correction to the classical action of the form 
\begin{widetext}
\bea
\label{gammaaction}
\Gamma^{(1)}[\langle \Psi\rangle,g, A]=\int d^4x\sqrt{-g}&& \left\{-\frac{\Lambda}{8\pi}+\frac{R}{16\pi}+\hbar(\alpha_1 R^2+\alpha_2 R_{ab}R^{ab}) -\frac{1}{4}F^{ab}F_{ab}  \right.\\
&& +k_1\left[a R F_{a b}F^{a b}+b R_{a b} F^{a c} F_{\hspace{0.15cm}c}^b+c R_{a b c d} F^{a b} F^{c d}+d\, \nabla^a F_{a b}\nabla_c F^{cb}+\frac{\hbar^2}{e^2}f(R^3)\right]\,  \nonumber\\
&& +k_2 \left[A (F_{ab}F^{ab})^2 + B F_{ab}F_{cd}F^{ac}F^{bd}+\dots \right]+\dots\, , \bigg\} \nonumber \, ,
\eea
\end{widetext}
where $a=5$, $b=-26$, $c=2$, $d=24$, $A=-5$, $B=14$, 
\bea
k_1=\frac{1}{2880\pi^2} \frac{\hbar e^2}{m_e^2}\, , \quad k_2=\frac{\hbar e^4}{2880\pi^2 m_e^4}\, , \label{k1k2constants}
\eea
and dots denote corrections with higher-order powers of the Riemann $R^a_{\,\,bcd}$ and field strength $F_{ab}$. The function $f(R^3)$ involves three powers of the Riemann tensor and will not be relevant. In  expression (\ref{gammaaction}), the coupling constant $\Lambda$ arises from absorbing the UV divergence of the zero-order term $E_0(x)$ by adding a suitable counterterm in the classical action (\ref{freeA}). This term physically represents  a cosmological constant. On the other hand, the UV divergence associated with the second-order term $E_2(x)$ can be reabsorbed in  Newton's gravitational constant. This yields some renormalized, observable value of $G$, which we have set to 1 according to our unit conventions. Finally, the UV divergences associated with the fourth-order term $E_4(x)$ can be reabsorbed in the fields $A_a$ and $F_{ab}$, as well as in some adimensional coupling constants $\alpha_1$, $\alpha_2$  by adding suitable counterterms in the original action. Their value have to be fixed with experimental measurements.

Higher-order terms in the heat kernel expansion ($E_{2k}(x)$ for $k\geq 3$) lead to finite  corrections to the original action in (\ref{1loopEA}). For example, the second line in equation (\ref{gammaaction}) displays the leading-order quantum corrections corresponding to $k=3$ in the heat kernel expansion, which were first computed by Drummond-Hathrell in \cite{Drummond:1979pp}. One can verify that each of these terms is of 6th adiabatic order. On the other hand, the Euler-Heisenberg perturbative corrections in (\ref{eq:Euler-HeisenbergLagrangian}) are of 8th adiabatic order, and are expected to arise from the $E_{8}(x)$ order in the heat kernel expansion. We have included them in the third line of (\ref{gammaaction}) for completeness.    For an explicit derivation of this coefficient, see e.g. Refs.~\cite{Avramidi:1990je,Avramidi:1990ug}.  

Notice that (\ref{gammaaction})  is independent of the vacuum state $\langle \Psi\rangle$; this information is  missed in the asymptotic expansion of the heat kernel (\ref{HKexpansion}), which is entirely constructed from the background fields $R_{abcd}$ and $F_{ab}$.

In this article we will explore new BH solutions of this effective action, and derive the master equations governing the propagation of linear gravitational and electromagnetic perturbations around them. The effective semiclassical Einstein and Maxwell equations are obtained by taking the field variations 
\bea
\label{variations}
\frac{\delta \Gamma^{(1)}[\langle \Psi\rangle, g,A]}{\delta g^{ab}}=0\, , \quad \frac{\delta \Gamma^{(1)}[\langle \Psi\rangle, g,A]}{\delta A^{a}}=0\, , 
\eea
respectively, which yield
\bea
G_{ab}=8\pi ( T^M_{ab}+\langle T^{(1)}_{ab} \rangle)\, , \quad \nabla_a F^{a}_{\,\,\, b}=\langle J^{(1)}_b \rangle\, , \label{fieldeqns}
\eea
for some  approximated expressions of $\langle T^{(1)}_{ab} \rangle$ and $\langle J^{(1)}_b \rangle$ obtained from the quantum corrections appearing in (\ref{gammaaction}). 

Our goal is to find static, spherically symmetric solutions of this system of equations with ADM mass $M$ and electric charge $Q$. Taking the full expression in (\ref{gammaaction}) can lead, however, to intractable equations, so we will assume some simplifications. First of all, the observed value of the cosmological constant $\Lambda$ is very small and it is only important at cosmological scales. Since we are interested in astrophysical local phenomena, we shall neglect it. Second, the coupling constants $\alpha_1$, $\alpha_2$ are severely constrained by observations, so in this work we shall set them to zero. On the other hand,  the term going with the prefactor $d$ will be of higher-order in $\hbar$ upon substituting back the result of solving the leading-order field equations, so it will not play a relevant role in what follows. Similarly, the term $f(R^3)$ is a purely gravitational correction (not included in Drummond-Hathrell original paper)  which is highly suppressed by Planck's constant and can be neglected in front of the rest of corrections. In conclusion, we can focus either on the $a$, $b$, $c$ corrections or on the $A$, $B$ corrections in (\ref{gammaaction}).

To further simplify the problem, we will consider the Drummond-Hathrell corrections ($a$, $b$, $c$ terms in (\ref{gammaaction})) separate from the Euler-Heisenberg $A$, $B$ terms. The latter dominate for higher electric fields, while the former are more important for BHs with low electric charge $Q$, like more realistic compact objects in astrophysics.

\section{A charged black hole with Drummond-Hathrell corrections}
\label{DHsolution}

In this section we solve the full semiclassical Einstein-Maxwell system of equations (\ref{fieldeqns}) with sources $\langle T_{ab} \rangle$, $\langle J^a \rangle$  determined by the Drummond-Hathrell corrections in the effective action (\ref{gammaaction}). We will focus  on static and spherically symmetric solutions, which are analogous to the classical Reissner-Nordstr\"om charged BH in general relativity. Then, we will study electromagnetic and  gravitational linear perturbations on this background. As mentioned above, the term multiplied by the coefficient $d$ in (\ref{gammaaction}) is of order $\mathcal{O}(\hbar)$ upon substituting back the result of solving the leading-order field equations, so we can safely ignore it. Therefore we focus on the $a$, $b$, $c$ corrections of  (\ref{gammaaction}).

Explicit expressions for $\langle T_{ab} \rangle$, $\langle J^a \rangle$ originated by these corrections  can be obtained straightforwardly by computing the functional derivatives (\ref{variations})  and using \texttt{xAct}. However, the results are cumbersome and not that illuminating, so we will omit them. 

Once we have explicit results for (\ref{fieldeqns}) we look for static and spherically symmetric solutions. The metric Ansatz for such a geometry is  
\be \label{metric0}
ds^2=N(r)\sigma(r) dt^2+\frac{dr^2}{N(r)}+r^2d\theta^2+r^2\sin^2\theta d\phi^2,
\ee
while the Ansatz for a  vector potential with similar properties reads 
\bea
A_a=\big(A_0(r),0,0,0\big). \label{A0}
\eea
Plugging these Ansatze on the semiclassical field equations (\ref{fieldeqns}) yields four independent, coupled differential equations. The Maxwell sector gives  one nontrivial equation (the $t$ component), while the gravitational sector produces only  three nontrivial and independent equations (the $tt$, $rr$ and $\theta\theta$ components). Again, these equations are still too tedious to show. To solve this system of coupled equations, we write
\begin{align}
N(r)&=\bigg(1-\frac{2M}{r}+\frac{Q^2}{r^2}\bigg)\big(1+k_1 \delta N(r)\big)\, ,\nonumber\\
\sigma(r)&=-1+k_1\delta \sigma(r)\, ,\nonumber\\
A_0(r)&=\frac{Q}{\sqrt{4\pi}r}+k_1\delta A(r)\, ,
\end{align}
for some unknown functions $\delta N(r)$, $\delta \sigma(r)$, $\delta A(r)$, where $k_1$ is given in (\ref{k1k2constants}). We have then three unknowns for four differential equations. If we substitute the above in the $tt$ and $rr$ components of the  semiclassical Einstein equations \eqref{fieldeqns}, and expand the result up to order $\mathcal{O}(k_1^2)$, we obtain
\bea
Q \left(-4 \sqrt{\pi } r^6 \delta A'+40 Q r (9 r-20 M)+304 Q^3+Q r^4 \delta\sigma\right)+r^5
   \left(r (r-2 M)+Q^2\right) \delta N'+r^4 \delta N (r^2-Q^2)&=&0,\quad \\
r^4 \left(r \left(-4 \sqrt{\pi } Q r \delta A'+\left(r^2-2 M r+Q^2\right)
   \left(\delta N'-\delta\sigma'\right)\right)+\delta N \left(r^2-Q^2\right)+Q^2
   \delta\sigma\right)+8 Q^2 \left(r (r-12 M)-6 Q^2\right)&=&0, \quad
\eea
while the only non-zero equation of the Maxwell sector, up to order $\mathcal{O}(k_1^2)$, gives 
\begin{align}
 4 \sqrt{\pi } r \left(r \delta A''+2 \delta A'\right)-Q \delta\sigma'-\frac{64}{r^5} \left(3
   M Q r+7 Q^3\right)=0.
\end{align}
 If we demand that the metric approaches Minkowski spacetime at spatial infinity (i.e. $\delta N(r)$, $\delta \sigma(r)$, $\delta A(r)\to 0$ as $r\to\infty$), and we fix the mass and charge of the resulting solution to be $M$ and $Q$, respectively, (by suitably identifying the $r^{-1}$ and $r^{-2}$ prefactors in the asymptotic expansion of the lapse function $N(r)\sigma(r)$)  the above system of equations produces
\beq
\label{eq:RNMetricCorrections}
\delta N(r)&=&\frac{5 Q^2 r\left(-184M+120 r\right)+416 Q^4}{5 r^4 \left(r (r-2M)+Q^2\right)},\nonumber\\
\delta \sigma(r)&=&\frac{88 Q^2}{r^4},\nonumber\\
\delta A(r)&=&\frac{2(3Q^3+10MQr)}{5\sqrt{\pi}r^5}\,.
\eeq
As a sanity check, we also verify that this solution satisfies the fourth independent  equation (the remaining $\theta\theta$ component of the semiclassical Einstein equations). 

Overall, the semiclassical metric and electromagnetic potential solutions are 
\bea
ds^2&=&-\left[1-\frac{2M}{r}+\frac{Q^2}{r^2}+k_1 \frac{8\left(-3 Q^{4}-5 M Q^{2} r+20 Q^{2} r^{2}\right)}{5 r^{6}} \right] dt^2+\frac{dr^2}{\left[1-\frac{2M}{r}+\frac{Q^2}{r^2}+k_1 \frac{416 Q^{4}-920 M Q^{2} r+600 Q^{2} r^{2}}{5 {r}^{6}} \right]} \nonumber\\
&& \quad+r^2 (d\theta^2+\sin^2\theta d\phi^2)\,, \label{DHg}\\
A(r)&=&\frac{Q}{\sqrt{4\pi}r}\left[1+k_2 \frac{4 \left(3 {Q}^{2}+10 {Mr}\right)}{5 r^{4}} \right]dt \label{DHA} \, .
\eea
Despite the presence of apparently different corrections in the $tt$ and $rr$ components of the spacetime metric, the (highest) roots of $g_{tt}(r_H)=0$ and $g^{-1}_{rr}(r_H)=0$ agree to give, to leading-order in the coupling constant $k_1$, 
\bea
r_H=r_+ + k_1 \frac{4r_- (11r_--35 r_+)}{5 r_+^2(r_+ - r_-)}+\mathcal O(k_1^2) \label{horizondh}\, ,
\eea
where $r_{\pm}=M\pm \sqrt{M^2-Q^2}$ are the  horizons of the classical Reissner-Nordstr\"om BH. Therefore, this new spacetime background still contains a BH horizon, located at $r=r_H$. 

Notice  that for $Q=0$ we recover the classical Schwarzschild geometry. In this case we also have $F_{ab}=0$, so indeed no quantum corrections are  expected in the metric to first order in $\hbar$, as argued in the Introduction.

\section{Perturbations on a neutral black hole with Drummond-Hathrell corrections}
\label{sec:partialproblem}

Our goal now is to derive the observable implications of these QED corrections, in such a way that they can be verified with future generation gravitational-wave detectors. To achieve this we need to study  electromagnetic and gravitational linear perturbations  on the new background solution, equations (\ref{DHg})-(\ref{DHA}). This problem is, however, rather complicated to address at first, because the resulting system of differential equations, derived from (\ref{fieldeqns}), couples both types of perturbations. As a first approach, in this section we will only study the propagation of  linear waves on our new background but for $Q=0$. In this case the electromagnetic potential (\ref{DHA}) vanishes entirely, and the spacetime metric (\ref{DHg}) reduces to the ordinary, neutral Schwarzschild geometry:
\beq  ds^2&=&-N dt^2+\frac{dr^2}{N}+r^2d\theta^2+r^2\sin^2\theta d\phi^2\,,\label{eq:SchwarzschildMetric} \eeq
where $N=1-\frac{2M}{r}$, with $M$ the BH mass.  This is an interesting problem in its own. Incidentally, we will correct some previous statements appearing in the literature. In the next section we will address the full problem, with $Q\neq 0$.

For $Q=0$    the semiclassical field equations (\ref{fieldeqns}) not only decouple, but the semiclassical Einstein equations become the ordinary classical ones. Therefore, the propagation of gravitational linear perturbations becomes trivial, i.e. one gets the well-known Regge-Wheeler and Zerilli equations for axial and polar perturbations, respectively\footnote{The RHS of the first equation in (\ref{fieldeqns}) is quadratic in $F$ for the Drummond-Hathrell corrections, so linear perturbations  around a background with  $F=0$  do not produce any deviation.}. The  problem reduces then to solving the one-loop corrected Maxwell's equations on a Schwarzschild background metric. A straightforward calculation yields 
\bea
\label{eq:Maxwellmodified}
M^{a}:=\nabla_b F^{(1)\, ab}-8k_1 R^{ab}_{\quad cd}\nabla_b F^{(1)\, cd}=0\, ,
\eea
for an electromagnetic perturbation $F^{(1)}_{ab}$ around a neutral background $F^{(0)}_{ab}=0$.\footnote{The RHS of the second equation in (\ref{fieldeqns}) is linear in $F$ for the Drummond-Hathrell corrections, so linear perturbations  around a background with  $F=0$  do  produce  deviations, as manifested in equation (\ref{eq:Maxwellmodified}).} In order to keep the notation consistent with the original paper by Drummond and Hathrell \cite{Drummond:1979pp},   in this section we will work with the coupling constant $\xi^2:=-8k_1$ for convenience.

Since the background metric is stationary and spherically symmetric, we expand the vector potential $A_a^{(1)}$ of the electromagnetic perturbation in Fourier modes of frequency $\omega$, as well as in vector spherical harmonics, labeled by  an angular momentum $\ell$ and   azimuthal number $m$ \cite{Cardoso:2001bb}: 
\begin{widetext}
  \begin{align}
    A^{(1)}_a(t,r,\theta,\phi) &= e^{-i\omega t}\sum_{\ell=1}^{\infty}\sum_{m=-\ell}^{\ell}
        \begin{pmatrix}
          \begin{bmatrix}
           0 \\           
           0 \\
           \frac{a^{\ell m}}{\sin\theta}\partial_\phi Y_{\ell m} \\
           -a^{\ell m}\sin\theta\partial_\theta Y_{\ell m}\\
          \end{bmatrix} +
          \begin{bmatrix}
           f^{\ell m}Y_{\ell m} \\
           h^{\ell m}Y_{\ell m} \\
           k^{\ell m}\partial_\theta Y_{\ell m} \\
           k^{\ell m}\partial_\phi Y_{\ell m} \\
         \end{bmatrix}
    \end{pmatrix}\,, \label{eq:vectorharmonics}
  \end{align}
\end{widetext}
where  $a^{\ell m}=a^{\ell m}(r)$ is a radial function. The first column vector represents the axial modes of parity $(-1)^{\ell+1}$ and the second column represents the polar modes of parity $(-1)^{\ell}$. Because the semiclassical Maxwell equations (\ref{eq:Maxwellmodified}) are still linear, axial and polar modes are expected to decouple from each other and can be analyzed independently. We will work  them out separately in the following two subsections.

\subsection{The axial sector}

We solve equation \eqref{eq:Maxwellmodified} for the axial modes using the substitution \eqref{eq:vectorharmonics}. The variable $a^{\ell m}$ is gauge invariant, since a gauge transformation $A^{(1)}_a \rightarrow A^{(1)}_a+\nabla_a\Lambda$
only affects the polar component of $A^{(1)}_a$ (in fact, the axial vector satisfies the Coulomb and Lorenz gauges identically). Doing this we obtain a second-order ordinary differential equation (ODE) that solves the full Maxwell equations\footnote{More precisely, eq. \eqref{eq:Maxwellmodified} with the Ansatz \eqref{eq:vectorharmonics} restricted to the axial modes gives only two nontrivial equations for $a^{\ell m}$, namely the $\theta$, $\phi$ components. Both are equivalent, as expected from spherical symmetry. Eq \eqref{eq:MEaxialquantum} is obtained from either one of them.}
\be
-N\frac{\ell(1+\ell)}{r^2}a^{\ell m}\left(4M\frac{\xi^2}{r^3}+1\right)+NN'\left(1-8M\frac{\xi^2}{r^3}+3\frac{\xi^2}{r^2}\right)\frac{d a^{\ell m}}{dr}
+\left(1-2M\frac{\xi^2}{r^3}\right)\bigg(N^2\frac{d^2 a^{\ell m}}{dr^2}+\omega^2a^{\ell m}\bigg)=0\,.\label{eq:MEaxialquantum}
\ee
We will rewrite this equation of motion for $a^{\ell m}$ in an explicit wavelike form. For this purpose, we further decompose $a^{\ell m}(r)=X_A(r)Z_A(r)$, for two arbitrary functions $X_A(r)$ and $Z_A(r)$. We fix the function $X_A$ by requiring that the quotient between the coefficients of $Z'_A$ and $Z''_A$ satisfy     $\frac{1}{N}\frac{dN}{dr}$, so that $Z_A(r)$ satisfies the usual ODE:
\begin{equation}
N^2\frac{d^2Z_A}{dr^2}+N N'\frac{dZ_A}{dr}+\bigg(\omega^2-V_A(r)\bigg)Z_A=0\,,\label{eq:Regge-Wheeler}
\end{equation}
for some potential function $V_A(r)$. Then, if we define the tortoise coordinate $r_*$ by the relation $dr_*/dr=1/N$,  Eq.~\eqref{eq:Regge-Wheeler} can  be expressed as a Regge-Wheeler equation
\begin{equation}\label{eq:Regge-WheelerA}
\frac{d^2Z_A}{dr_*^2}+\bigg(\omega^2-V_A\bigg)Z_A=0\,.
\end{equation}
Doing all the above we find, 
\begin{equation}
X_A(r)=\frac{r^{3/2}}{\sqrt{-2 \xi^2 M + r^3}}\, ,
\end{equation}
and
\beq
\label{eq:AxialPotentialQuantum}
V_A&=&\frac{N}{\Upsilon} \bigg[\frac{L}{r^2}+2\frac{\xi^2M}{r^6}(15M+r(L-6))-\frac{\xi^4M^2}{r^9}(42M+r(8L-15)) \bigg]\,,\\
&=& N\frac{L}{r^2}+N\frac{6\xi^2M(5M+(L-2)r)}{r^6}+\mathcal{O}\left(\xi^4\right)\,,\nonumber
\eeq
where we defined $L=\ell(\ell+1)$ and $\Upsilon=\left(1-\frac{2M\xi^2}{r^3}\right)^2$. In summary, we can cast the semiclassical  Maxwell equation for axial waves as an ordinary Regge-Wheeler equation for $Z_A(r)$, where all quantum corrections are encoded in the effective potential (\ref{eq:AxialPotentialQuantum}).

\subsection{The polar sector}

A similar analysis can be done for the polar sector, although the calculation is  more involved in this case. Maxwell equations \eqref{eq:Maxwellmodified} for the polar sector of \eqref{eq:vectorharmonics} yield three independent  equations that couple the three functions $f^{\ell m}$, $h^{\ell m}$, $k^{\ell m}$. On top of that, these functions are gauge dependent. To work with physically relevant, gauge-independent variables we have to take suitable linear combinations of these three functions. These can be inferred from the different components of the field strength $F_{ab}=\nabla_a A_b - \nabla_b A_a$. For instance, the $tr$ component motivates us to define
\bea
\label{eq:PsiSphericalHarmonics}
\psi_{\ell m}(r)=\bigg[-i\omega h^{\ell m}-\frac{\partial f^{\ell m}}{\partial r}\bigg] r^2\, .
\eea
It is easy to check that this combination remains invariant under a gauge transformation $A_a \to A_a+\nabla_a\Lambda$, for any $\Lambda$. The other two gauge-invariant combinations that one can build from the  functions $f_{\ell m}$, $h_{\ell m}$, $k_{\ell m}$ can  be deduced from the $t\theta$ and $r\theta$ components of $F_{ab}$: $\psi^{(1)}_{\ell m}=f^{\ell m}+i \omega k^{\ell m}$, $\psi^{(2)}_{\ell m}=h^{\ell m}-\partial_r k^{\ell m}$. The change $\{f^{\ell m},h^{\ell m},k^{\ell m}\}\to \{\psi_{\ell m}, \psi^{(1)}_{\ell m}, \psi^{(2)}_{\ell m}\}$ drastically simplifies  the  set of Maxwell equations (\ref{eq:Maxwellmodified}). Namely, the $t$ component allows us to solve $\psi_{\ell m}^{(1)}$ in terms of $\psi_{\ell m}$ and its derivative:
\be
\psi^{(1)}_{\ell m}=\frac{(2M-r)(-12M\xi^2 \psi_{\ell m}+r(4M\xi^2+r^3)\partial_r \psi_{\ell m})}{(r^3-2M\xi^2)r^2\ell(\ell+1)},
\ee
while the $r$ component solves $\psi_{\ell m}^{(2)}$ in terms of $\psi_{\ell m}$ alone:
\bea
\psi^{(2)}_{\ell m}=\frac{r(4M\xi^2+r^3)(-i \omega \psi_{\ell m})}{(r^3-2M\xi^2)(2M-r)\ell(\ell+1)} \, .
\eea
 The $\theta$ (or $\phi$) component of the semiclassical Maxwell equations (\ref{eq:Maxwellmodified}) becomes trivial using these results. The polar sector is then reduced to solving for one single, gauge-independent variable $\psi_{\ell m}(r)$.

To find the differential equation governing the radial profile of $\psi_{\ell m}(r)$, we need to manipulate  the three independent Maxwell equations so as to get one differential equation that only involves the $h^{\ell m}$ and $f^{\ell m}$ functions. One can check that the combination of Maxwell equations $\partial_r(r^2 \frac{N}{\sqrt{\Upsilon}} M^t)+\partial_t(\frac{r^2}{N\sqrt{\Upsilon}}M^r)=0$, where $M^a$ was defined in (\ref{eq:Maxwellmodified}), accomplishes this. Moreover, it only involves the combination \eqref{eq:PsiSphericalHarmonics}: 
\begin{align}
    &rN\bigg[96M^3\xi^4+4(12+L)M\xi^2r^4-Lr^7-4M^2\xi^2r(30r^2+(6+L)\xi^2)\bigg]\psi_{\ell m}\nonumber\\
    & +2M r^2N (-32M^2\xi^4+12M\xi^4r+32M\xi^2r^3-15\xi^2r^4+r^6)\frac{d\psi_{\ell m}}{d r}\nonumber\\
    &-r^4N^2(2M\xi^2-r^3)(4M\xi^2+r^3)\frac{d^2\psi_{\ell m}}{dr^2}-\omega^2r^4(2M\xi^2-r^3)(4M\xi^2+r^3)\psi_{\ell m}=0.
\end{align}
This is a second-order ODE for $\psi_{\ell m}(r)$. Now, as in the axial case, we decompose  $\psi_{\ell m}(r)=X_P(r)Z_P(r)$ and demand that $Z_P$ obeys the Regge-Wheeler-like equation~\eqref{eq:Regge-Wheeler}:
\begin{equation}
N^2\frac{d^2Z_P}{dr^2}+N N'\frac{dZ_P}{dr}+\bigg(\omega^2-V_P(r)\bigg)Z_P=0\,.\label{eq:Regge-WheelerP}
\end{equation}
As a result we find 
\begin{equation}
X_P(r)=\frac{r^{3/2}\sqrt{-2M\xi^2+r^3}}{4M\xi^2+r^3}\,,
\end{equation}
and the potential
\beq
\label{eq:PolarPotentialQuantum}
V_P&=&\frac{N}{\Upsilon \Upsilon_2}\bigg[\frac{L}{r^2}-\frac{6\xi^2M(5M+r(L-2))}{r^{6}}+\frac{3\xi^4M^2(r(17+4L) - 38M)}{r^{9}}+\frac{4\xi^6M^3(6M+r(3-2L))}{r^{12}}\bigg]\\
&=&N\frac{L}{r^2}-N\frac{6M\xi^2[5M+(L-2)r]}{r^6}+\mathcal{O}\left(\xi^4\right)\,,\nonumber
\eeq
with $\Upsilon_2=1+4\xi^2M/r^3$.  Again, all quantum corrections are encoded in the effective potential.

Notice  that, already at leading-order, the two potentials (\ref{eq:AxialPotentialQuantum}) and (\ref{eq:PolarPotentialQuantum}) differ (the coefficient of $\xi^2$ has different sign in both potentials). Therefore, the axial and polar sectors  are not isospectral and we do not have superpartner potentials (as can be easily verified). We  calculate the QNM frequency spectra in the following subsection, and verify that  isospectrality is indeed broken by vacuum fluctuations. In essence, this is because the quantum correction in (\ref{fieldeqns}) can be interpreted as an effective current $\langle J^a\rangle$ that breaks the classical electric-magnetic duality symmetry of the vacuum Maxwell equations. If we regard the axial modes as the electric component of the electromagnetic perturbation (odd under parity transformations) and the polar component as the magnetic component (even under parity transformations), the lack of isospectrality we find is a direct consequence of explicitly breaking this symmetry.

Incidentally, Eq. (\ref{eq:Maxwellmodified}) agrees with the field equation that one obtains in a generalized electromagnetic theory ~\cite{Chen:2013ysa}, where a  coupling between the field $F_{ab}$ and the Weyl tensor is studied. Our results for the axial and polar potentials match those obtained in Ref.~\cite{Chen:2013ysa} with the identification $\xi^2=4\alpha$, where $\alpha$ is the relevant coupling constant in Ref.~\cite{Chen:2013ysa}. Despite this agreement, the authors of Ref.~\cite{Chen:2013ysa} claim to have obtained superpartner potentials, but their own  results, which also indicate a different spectrum for axial and polar perturbations,  disprove these claims.

\subsection{Quasinormal mode spectrum} \label{qnmsection}
A BH can be perturbed in  various ways, and  small perturbations typically satisfy a wave equation like in \eqref{eq:Regge-WheelerA} or \eqref{eq:Regge-WheelerP}. The   QNMs of a BH are the proper solutions of these perturbation equations. In other words, they are solutions that satisfy boundary conditions appropriate for a dissipative system: purely ingoing waves at the BH horizon and purely outgoing waves at spatial infinity. These solutions only exist for certain characteristic complex-valued frequencies $\omega$, called quasinormal frequencies, whose real and imaginary parts represent the oscillatory frequency and decaying timescale of the propagating linear (scattered) field~\cite{Kokkotas:1999bd,Nollert_1999,Berti:2009kk,Konoplya:2011qq}. These characteristic oscillations always appear and dominate the signal at intermediate times in any event involving BHs. For example, they can appear at the linearized level, where fields are treated as a perturbation in a single BH spacetime (as in the present article), but also in full numerical simulations of BH-BH collisions or stellar collapse, making them a central feature for tests involving gravitational waves. 

We wish to obtain now the characteristic frequencies of QNMs associated with the axial and polar wave equations described by the effective potentials \eqref{eq:AxialPotentialQuantum} and \eqref{eq:PolarPotentialQuantum}. Despite the fact that QNMs are triggered by external perturbations, their frequencies constitute an intrinsic property of the background, and therefore encode crucial geometric information about BHs. Moreover, QNM overtones have been proposed as a possible probe into the quantum aspects of spacetime (see \cite{Jaramillo:2020tuu} and references therein). 

The method we implement here makes use of a geometric frame based on conformal compactifications, together with hyperboloidal foliations of spacetime. Methodologically, a compactified hyperboloidal approach to QNMs is adopted to cast QNMs in terms of the spectral problem of a non-self-adjoint operator. Crucially, such a spectral problem can be cast as a proper “eigenvalue problem” for this non-self-adjoint operator. Therefore, following \cite{Macedo_2018}, \cite{Macedo_2020}, we construct numerically the pseudospectrum notion via Chebyshev spectral methods. \footnote{This is also relevant for addressing the potential spectral instability of a class of non-self-adjoint operators, which are associated with a nonconservative system like in a BH, where field perturbations leak away from the system at far distances and through the BH horizon.} Since the actual value of $\xi^2$ is way below the machine precision, our strategy will be to do the calculation for several artificially big values of $\xi^2$, and then to extract the actual numerical value using a linear extrapolation.

\begin{table}
\begin{tabular}{|c|c||c|c|}
 \hline
$\ell$&  $\xi/M$ &$M \text{Re}(\omega_{0\ell})$ &$M \text{Im}(\omega_{0\ell})$\\
 \hline
\,1\, &$0.00$&0.24826&$-0.092486$\\
 1& $0.07$&0.24842&$-0.092477$\\
 1&$0.09$&0.24852&$-0.092473$\\
1& $0.10$&0.24858&$-0.092468$\\
1&  $0.15$&0.24898&$-0.092447$\\
1& $0.20$&0.24953&$-0.092425$\\ \hline
2& $0.00$&0.45760&$-0.095004$\\
2&  $0.07$&0.45785&$-0.095000$\\
2& $0.09$&0.45802&$-0.094997$\\
2& $0.10$&0.45812&$-0.094996$\\
 2& $0.15$&0.45878&$-0.094986$\\
 2& $0.20$&0.45970&$-0.094976$\\
 \hline
\end{tabular}
\caption{Fundamental mode of axial electromagnetic QNMs (see Eq.~\eqref{eq:Regge-WheelerA}) for different values of the coupling constant $\xi$ and angular dependence $\ell$. Columns show the real and imaginary parts (normalized by BH mass $M$) of the mode.}\label{table:FundamentalModeAxial}
\end{table}
\begin{table}
\begin{tabular}{|c|c||c|c|}
 \hline
 $\ell$&  $\xi/M$ &$M \text{Re}(\omega_{0\ell})$ &$M \text{Im}(\omega_{0\ell})$\\
 \hline
1& $0.00$&0.24826&$-0.09249$\\
 1& $0.07$&0.24810&$-0.09250$\\
 1&$0.09$&0.24801&$-0.092505$\\
 1&$0.10$&0.24794&$-0.09251$\\
  1&$0.15$&0.24755&$-0.09253$\\
 1&$0.20$&0.24700&$-0.09257$\\ \hline
2& $0.00$&0.45760&$-0.095004$\\
 2& $0.07$&0.45734&$-0.095009$\\
 2&$0.09$&0.45717&$-0.095012$\\
 2&$0.10$&0.45707&$-0.095014$\\
 2& $0.15$&0.45642&$-0.095028$\\
 2& $0.20$&0.45550&$-0.095051$\\
 \hline
\end{tabular}
\caption{Fundamental mode of polar electromagnetic QNMs (see Eq.~\eqref{eq:Regge-WheelerP}) for different values of the coupling constant $\xi$ and angular dependence $\ell$. Columns show the real and imaginary parts (normalized by BH mass $M$) of the mode.}\label{table:FundamentalModePolar}
\end{table}

Specific results are listed in the Tables \ref{table:FundamentalModeAxial} and \ref{table:FundamentalModePolar}\footnote{The lower bound $\xi=0.07M$ is the minimum value of $\xi$ that accomplishes to see the impact of quantum corrections within 5-6 digits of reliable precision. For smaller values of $\xi$ one would need more digits of precision to find significant deviations from the classical values $\omega_0$. On the other hand, the upper bound $\xi=0.2M$ represents  the  value of $\xi$ beyond which the linear truncation  starts to fail.}. We have also checked these results with the more familiar direct integration method~\cite{Berti:2009kk} (see Appendix A for a review). Using these results one can now show that the difference between the classical and semiclassical predictions  scales linearly with $\xi^2\sim \hbar e^2/m_e^2$. This is clearly shown in FIGs. \ref{axialw} and \ref{polarw}. If for each multipole $\ell$ we define
\be
\delta_\ell\equiv \left(\text{Re}(\omega-\omega_0),\text{Im}(\omega-\omega_0)\right),
\ee
where $\omega_0$ is the classical QNM value (i.e. the $\xi=0$ value), then one can find 
\beq
&&\delta_1^\text{axial}=-(0.0317,0.0017)\frac{\hbar e^2}{360\pi^2 m_e^2 M^3},\,\label{axial1}\\
&&\delta_2^\text{axial}=-(0.0526, 0.0008)\frac{\hbar e^2}{360\pi^2 m_e^2 M^3},\, \\
&&\delta_1^\text{polar}=(0.0315,0.0020)\frac{\hbar e^2}{360\pi^2 m_e^2 M^3},\\
&&\delta_2^\text{polar}=(0.0524, 0.0012)\frac{\hbar e^2}{360\pi^2 m_e^2 M^3}\,. \label{polar2}
\eeq
Notice the opposite sign in the correction of axial and polar modes, showing immediately the breaking of isospectrality. 

These results can be checked against model-independent expansions of the relevant effective potentials~\cite{Cardoso:2019mqo,McManus:2019ulj}. More precisely, for $\ell=1$ the effective potentials (\ref{eq:AxialPotentialQuantum})-(\ref{eq:PolarPotentialQuantum}) yield
\be
V_{A,P}/N=\frac{\ell(\ell+1)}{r^2}\pm \frac{30\xi^2M^2}{r^6}\,,
\ee
which in the terminology of Refs~\cite{Cardoso:2019mqo,McManus:2019ulj} amounts to having $\delta V=1/r_H^2(r_H/r)^6\beta_6^1$, with $\beta_6^1=\pm 15\xi^2/(8M^2)$. From tabulated values in those papers one can find that $M\delta \omega=\pm(0.0317, 0.00184)\xi^2/M^2$, in very good agreement with the above results. On the other hand, for $\ell=2$  equations (\ref{eq:AxialPotentialQuantum})-(\ref{eq:PolarPotentialQuantum}) produce
\be
V_{A,P}/N=\frac{\ell(\ell+1)}{r^2}\pm \left( \frac{30\xi^2M^2}{r^6}+ \frac{24M \xi^2}{r^5} \right)\,,
\ee
From Equations (7)-(8) in~\cite{Cardoso:2019mqo} we infer  $\delta V=1/r_H^2(r_H/r)^5\beta_5^1+1/r_H^2(r_H/r)^6\beta_6^1$, with $\beta_5^1=\pm 24\xi^2/(8M^2)$ and $\beta_6^1=\pm 15\xi^2/(8M^2)$. Then, Equation (11) in~\cite{Cardoso:2019mqo} using tabulated values leads to $M\delta \omega=\pm(0.0525, 0.0009)\xi^2/M^2$, which is again in good agreement with our results above.

\begin{figure}[t!]
\begin{center}
\includegraphics[width=0.49\textwidth]{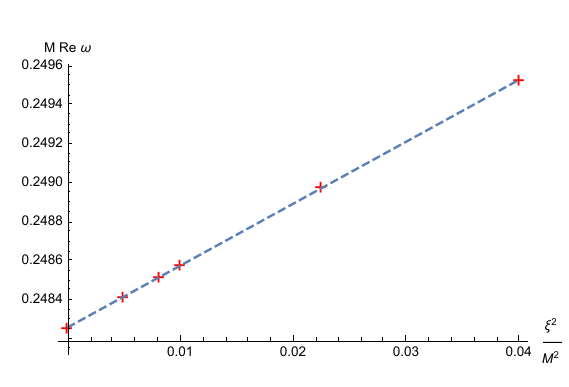}
\includegraphics[width=0.49\textwidth]{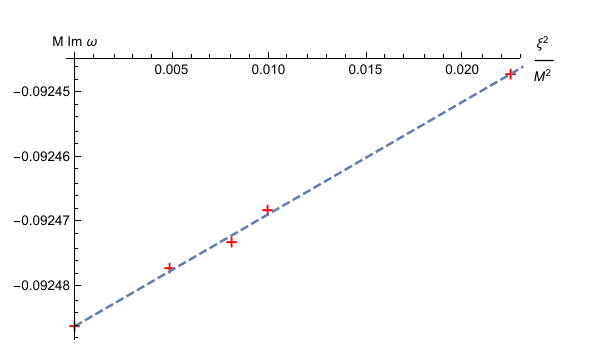}
\includegraphics[width=0.49\textwidth]{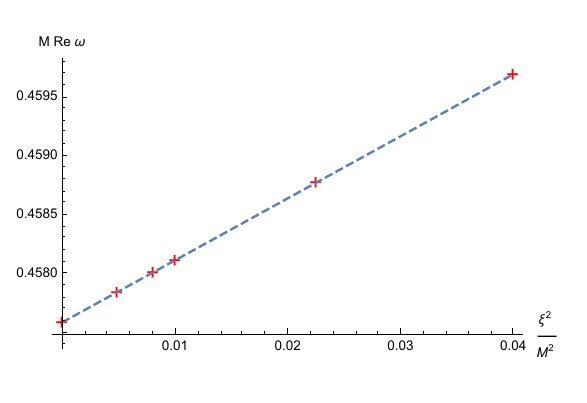}
\includegraphics[width=0.49\textwidth]{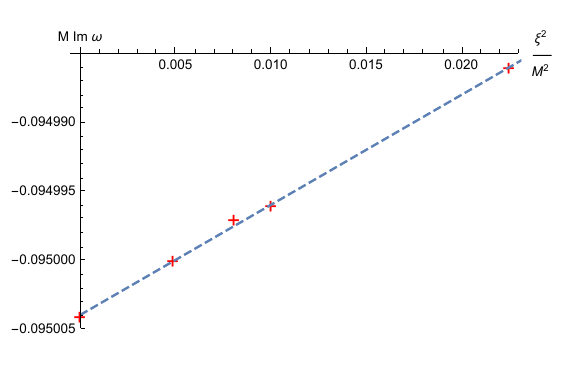}
\caption{Linear dependence of the fundamental QNM frequency $\omega_{0\ell}$ with Planck's constant $\xi^2/M^2$ for $\ell=1$ (upper) and $\ell=2$ (lower) in case of axial electromagnetic perturbations. Red crosses denote the numerically calculated quasinormal frequencies, while the blue dashed line represents a linear fit.}
\label{axialw}
\end{center}
\end{figure}

\begin{figure}[t!]
\begin{center}
\includegraphics[width=0.49\textwidth]{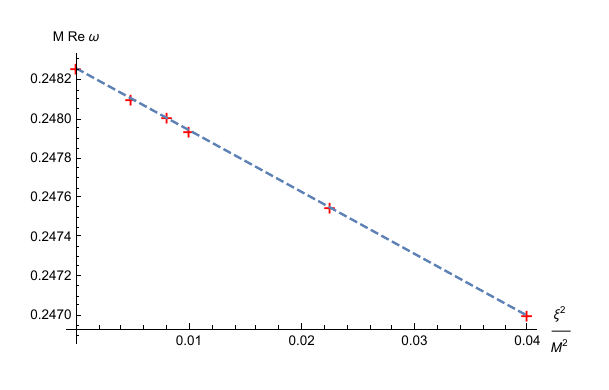}
\includegraphics[width=0.49\textwidth]{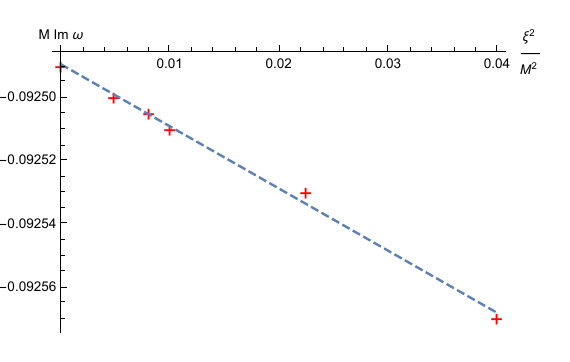}
\includegraphics[width=0.49\textwidth]{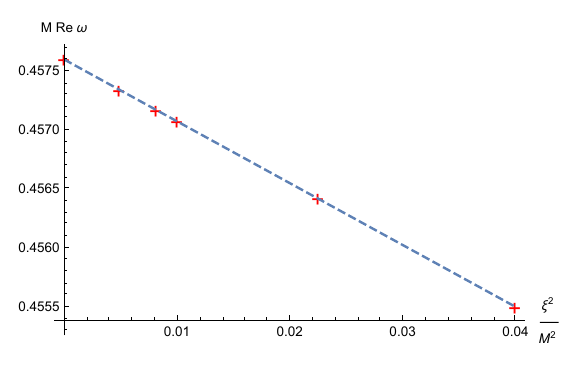}
\includegraphics[width=0.49\textwidth]{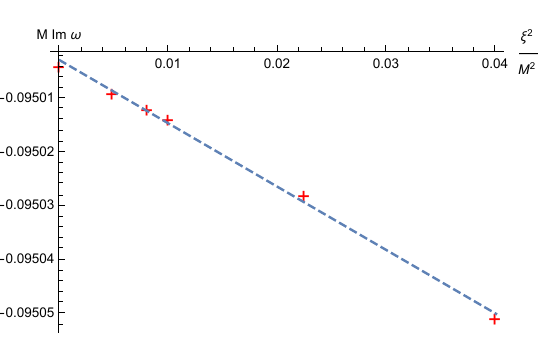}
\caption{Linear dependence of the fundamental QNM frequency $\omega_{0\ell}$ with Planck's constant $\xi^2/M^2$ for $\ell=1$ (upper) and $\ell=2$ (lower) in case of polar electromagnetic perturbations. Red crosses denote the numerically calculated quasinormal frequencies, while the blue dashed line represents a linear fit.}
\label{polarw}
\end{center}
\end{figure}

\subsection{The static limit}

 In the classical theory,  the only static  solution of Maxwell equations in a Schwarzschild BH spacetime, which is additionally spherically symmetric and vanishes at spatial infinity, is  the field of a point  electric  charge (i.e. the monopole $\ell=0$ solution). This  is precisely  the electromagnetic potential of the Reissner-Nordstr\"om BH solution. Therefore, if we add this static electromagnetic field as a small perturbation on the Schwarzschild BH background, one can roughly say that the spacetime becomes a Reissner-Nordstrom BH after ``eating'' this electric charge. 
 
 Similarly,  the  static electrically charged BH  solution of the full semiclassical field equations (\ref{fieldeqns}) that we obtained in Sec. \ref{DHsolution}, Eqs. (\ref{DHg})-(\ref{DHA}), should be compatible with a Schwarzschild BH that ``eats'' a static solution of the semiclassical  Maxwell equations  \eqref{eq:Maxwellmodified}. In other words, the static solutions ($\omega=0$) of (\ref{eq:Regge-Wheeler}) and (\ref{eq:Regge-WheelerP}) should be compatible, or partially recover, the electromagnetic potential obtained in (\ref{eq:RNMetricCorrections}).

Let us make this idea more precise.
In the static limit ($\omega=0$), the master equations (\ref{eq:Regge-Wheeler}) and (\ref{eq:Regge-WheelerP}) reduce to
\be
\label{eq:StaticLimit}
\left(N\Psi_{\ell}'\right)'-\frac{V_{\ell}}{N}\Psi_{\ell}=0\,,
\ee
where a prime stands for radial derivative. At infinity $V_{\ell}\sim \ell(\ell+1)/r^2$ while $N\sim 1$, thus solutions behave as $\Psi_{\ell} \sim a_1r^{-\ell}+a_2 r^{\ell+1}$ for some constants $a_1$, $a_2$. At the horizon, since $V_{\ell}/N$ is finite, we find $\Psi_{\ell}\sim c_1+c_2\log(r-2M)$, for some constants $c_1$, $c_2$. Demanding regularity at the horizon requires $c_2=0$, while regularity at infinity demands $a_2=0$ for any $\ell$. 
However, this is a big constraint, since fixing $c_2=0$ as a boundary condition will in general correspond to $a_1\neq 0$, $a_2\neq 0$ at infinity.

To see the  implications of imposing these regularity conditions, let us first analyze the classical case. First of all, multiply  \eqref{eq:StaticLimit} by the complex conjugate $\Psi_{\ell}^*$ and integrate outside the horizon to find,
\beq
&0&=\int_{2M}^\infty dr\left( \left(N\Psi'_{\ell}\right)'\Psi_{\ell}^*-\frac{V_{\ell}}{N}\left|\Psi_{\ell}\right|^2\right)=\bigg[N\Psi'_{\ell}\Psi_{\ell}^*\bigg]_{2M}^\infty-\int_{2M}^\infty dr\left( \left(N\left|\Psi'_{\ell}\right|^2\right)+\frac{V_{\ell}}{N}\left|\Psi_{\ell}\right|^2\right)\,.
\eeq
The first term is zero for any $\ell$ as a consequence of the regularity conditions imposed above. On the other hand, for positive definite potentials, as in the classical case, the integral in the second term is definite positive. If $\ell\neq 0$ then $V_{\ell}\neq 0$ and the above equation implies $|\Psi_{\ell}|=0$ and $|\Psi'_{\ell}|=0$, i.e. the only regular solution is  $\Psi_{\ell}=0$. If $\ell=0$, in the classical case we would have $V_0=0$, and the above equation would only imply $|\Psi'_{\ell}|=0$ so the most general solution is $\Psi_0=\eta=$const. The axial sector is trivial for $\ell=0$, since the angular derivatives vanish. This solution is necessarily polar. Using then \eqref{eq:PsiSphericalHarmonics} this result implies $f^{\ell m}=Q/r\,\delta_{\ell0}\delta_{m0}$, which, according to (\ref{eq:vectorharmonics}), is the electromagnetic potential of a point charge. With this static electromagnetic perturbation, the fixed Schwarzschild BH background becomes, in a natural way,  a Reissner-Nordstr\"om charged BH. This is the expected result.

However, quantum corrections make the potential $V_{\ell}$ not positive definite, so the above argument fails and there could be nontrivial solutions satisfying the regularity conditions at both the horizon and infinity. Let us look for solutions of the form
\be
\label{eq:PsiExpansion}
\Psi_0(r)=\eta+\xi^2\chi(r),
\ee
where $\eta$ is the constant solution satisfying the classical static limit equation for $\ell=0$ and $\chi$ is the leading-order quantum correction. Upon substituting this expansion in equation \eqref{eq:StaticLimit}, with the potential given by \eqref{eq:PolarPotentialQuantum} and letting $\ell=0$, we have that the solution of the master equation is 
\be
\chi(r)=-\frac{M\eta}{r^3}+c_1-c_2\left(r+2\log{(r-2M)}\right).
\ee
By demanding regularity at the horizon and at spatial infinity, then $\chi(r)=c_1-M\eta/r^3$. Therefore, 
\be
\Psi_0\sim\eta-\xi^2M\eta/r^3+c_1\xi^2. \label{l0mode}
\ee
This is the $\ell=0$ mode. For general $\ell\neq 0$, we can do a similar reasoning:
\be
\label{eq:PsiExpansion}
\Psi_{\ell}(r)=0+\xi^2\chi_{\ell}(r),
\ee
The solution to the Regge-Wheeler equation restricted to order $\xi^2$ in this case is given in terms of special functions. Demanding regularity conditions at both the horizon and infinity eventually renders $\chi_{\ell}=0$ for any $\ell\neq 0$. 

Eq. (\ref{l0mode}) is polar, so from equation \eqref{eq:PsiSphericalHarmonics}, we obtain that $f^{\ell m}\sim c_1/r+\eta M \xi^2/r^4$. Recalling that $f^{\ell m}$ is the $t$-component of $A_a$ in (\ref{eq:vectorharmonics}), this finding is consistent with equation \eqref{eq:RNMetricCorrections}  ($c_1$ can be reabsorbed in $Q$, and $\eta\propto Q$). In particular, this  is the leading-order result for  \eqref{eq:RNMetricCorrections}   for small charge $Q$.

\section{Axial perturbations on a charged black hole with Drummond-Hathrell corrections}
\label{sec:DH}

In this section we address the original problem posed at the beginning of Sec. \ref{sec:partialproblem}. Namely, we will derive the coupled system of equations governing the propagation of both electromagnetic and gravitational linear perturbations on the background BH solution  (\ref{DHg})-(\ref{DHA}), with $Q\neq 0$. Still, we will only deal with the axial case, which is simpler. The complexity of the problem is already formidable, and all calculations require using the software \texttt{xAct} for tensor algebra manipulations with {\it Mathematica}. The polar case, technically much more involved, is  qualitatively similar and is not expected to provide any new insight.

Using (\ref{gammaaction}), and restricting to the $a$, $b$, $c$ corrections, we derive the explicit form of the semiclassical field equations (\ref{fieldeqns}). Then, we look for solutions of these equations with a metric and electromagnetic potential of the form 
\bea \label{decomp}
{g}_{ab}&=&g_{ab}^{(0)}+g^{(1)}_{ab},\nonumber\\
{A}_{a}&=&A_{a}^{(0)}+A^{(1)}_{a},
\eea
where $g_{ab}^{(0)}$ and $A_{a}^{(0)}$ are given in Eqs. (\ref{DHg})-(\ref{DHA}) of Sec. \ref{DHsolution}. For the electromagnetic perturbation we expand again in Fourier modes of frequency $\omega$ and vector spherical harmonics of odd parity
  \begin{align}
    A^{(1)}_a(t,r,\theta,\phi) &= e^{-i\omega t}\sum_{\ell=1}^{\infty}\sum_{m=-\ell}^{\ell}
                \begin{bmatrix}
           0 \\           
           0 \\
           \frac{a^{\ell m}(r)}{\sin\theta}\partial_\phi Y_{\ell m} \\
           -a^{\ell m}(r)\sin\theta\partial_\theta Y_{\ell m}\\
          \end{bmatrix} \,, \label{DHa}
  \end{align}
for some radial functions $a^{\ell m}(r)$.  For the gravitational perturbation we have to expand  in terms of  tensor spherical harmonics of odd parity. In the Regge-Wheeler gauge fixing this can be written as \cite{Berti:2009kk}
\begin{widetext}
  \begin{align}
    g^{(1)}_{ab}(t,r,\theta,\phi) &= e^{-i\omega t}\sum_{\ell=2}^{\infty}\sum_{m=-\ell}^{\ell}
\left[\begin{array}{cccc}
0 & 0 & -h_0(r)\csc\theta\frac{\partial Y_{\ell m}}{\partial\theta} & h_0(r)\sin\theta\frac{\partial Y_{\ell m}}{\partial\theta} 	\\
0 & 0 & -h_1(r)\csc\theta\frac{\partial Y_{\ell m}}{\partial\phi} & h_1(r)\sin\theta\frac{\partial Y_{\ell m}}{\partial\theta}	\\
-h_0(r)\csc\theta\frac{\partial Y_{\ell m}}{\partial\phi} & -h_1(r)\csc\theta\frac{\partial Y_{\ell m}}{\partial\phi} & 0 & 0	\\
h_0(r)\sin\theta\frac{\partial Y_{\ell m}}{\partial\phi} & h_1(r)\sin\theta\frac{\partial Y_{\ell m}}{\partial\theta} & 0 & 0
\end{array}\right]\, ,
 \label{DHh}
  \end{align}
\end{widetext}
for some radial functions $h_{0,\ell m}(r)$, $h_{1,\ell m}(r)$. Then, the problem reduces  to determining the three unknown functions $h_{0,\ell m}(r)$, $h_{1,\ell m}(r)$, $a^{\ell m}(r)$ by perturbing the  semiclassical field equations to first order in perturbations:
\bea
E_{ab}&:=&\delta G_{ab}-8\pi\delta\left(T_{ab}^M+\langle T_{ab}\rangle\right)=0, \label{perturbE}\\
M^{b}&:=&   \nabla_a \delta F^{ab}-\delta\langle J^b\rangle=0, \label{perturbM}
\eea
taking into account that $\delta g_{ab}=g^{(1)}_{ab}$ and $\delta A_{a}=A_{a}^{(1)}$.
Explicit expressions for the linearized semiclassical  equations can be obtained using \texttt{xAct}, but the output is too cumbersome to fit in these pages.

\subsection{Gravitational perturbations}

By introducing (\ref{DHa}) and (\ref{DHh}) in the perturbed semiclassical Einstein equation, (\ref{perturbE}), we get four independent, nontrivial equations: $E_{t\theta}=0$, $E_{r\theta}=0$, $E_{\theta\theta}=0$, $E_{\theta\phi}=0$. The equation $E_{\theta\phi}=0$ can be solved for $h^{\ell m}_0(r)$ in terms of $h^{\ell m}_1(r)$ and its derivatives.  Then $E_{\theta\theta}=0$ is satisfied identically, and we remain with two independent equations, $E_{t\theta}$ and $E_{r\theta}$, and one unknown function $h^{\ell m}_1(r)$. Doing the transformation $h^{\ell m}_1(r)=Q^{\ell m}_{\rm axial}(r)\frac{r}{1-\frac{2M}{r}+\frac{Q^2}{r^2}}$, the equation $E_{r\theta}=0$ leads to a second-order differential equation for $Q^{\ell m}_{\rm axial}(r)$:
\bea
\left[C_1(r) \frac{d^2}{dr^2}+ C_2(r) \frac{d}{dr}+ (\omega^2- C_3(\omega,\ell,r) ) \right]Q^{\ell m}_{\rm axial}=C_4(\omega, r)a^{\ell m} + C_5(\omega, r)\frac{da^{\ell m}}{dr}  \label{DHE1}\, ,
\eea
where 
\bea
C_1 &=& \frac{\left(Q^{2}+r(-2 M+r)\right)^{2}}{r^{4}}+\frac{8 k_1 Q^{2}\left(Q^{2}+r(-2M+r)\right)\left(179 Q^{2}+5 r(-76 M+45 r)\right)}{5 r^{8}} \, ,\nonumber\\
C_2 &=& -\frac{2\left(Q^{2}-M r\right)\left(Q^{2}+r(-2 M+r)\right)}{r^{5}}-\frac{8 k_1 Q^{2}\left(812 Q^{4}+Q^{2} r(-3033 M+1496 r)+5 r^{2}\left(568 M^{2}-569 M r+146 r^{2}\right)\right)}{5 r^{9}}\, ,\nonumber \\
C_3 &=& \frac{\left(Q^{2}+r(-2 M+r)\right)\left(4 Q^{2}+r(-6 M+L r)\right)}{r^{6}} \nonumber\\
&-&   \frac{ 208 k_1 Q^{2}\omega^{2}}{r^4 }\nonumber\\
&-& 8 k_1 Q^{2}\left[\frac{1152 Q^{6}-Q^{4} r(6831 M+(-3392+47 L) r)+Q^{2} r^{2}\left(13632 M^{2}+(-13768+199 L) M r-3(-1186+39 L) r^{2}\right)}{(Q^{2}+r(-2 M+r))5 r^{10}}\right]\nonumber\\
&-& 8 k_1 Q^{2} \frac{\left[-1840 M^{3}-42\left(-68+L\right) M^{2} r+(-1525+49 L) M r^{2}-2(-141+7 L) r^{3}\right]}{(Q^{2}+r(-2 M+r)) r^{7}} \, ,\nonumber\\
C_4 &=& \frac{8 i \sqrt{\pi} Q \omega(Q^{2}+r(-2 M+r))}{r^5}+\frac{8 i k_1 \sqrt{\pi} Q\omega(Q^{2}+r(-2 M+r))(480 Q^{2}+312 r(-2M+r))}{r^9}\, ,\\
C_5 &=& -\frac{1056  { i k_1 } \sqrt{\pi} {Q \omega}\left({Q}^{2}+{r}(-2 {M}+{r})\right)^{2}}{{r}^{8}} \, ,
\eea
and $L=\ell(\ell+1)$.  To arrive at this expression we have made use of the classical result to get rid of a source term $k_1 \frac{d^2a^{\ell m}}{dr^2}$, which is allowed to first order in $k_1$. In the classical limit $k_1=0$ our result recovers the expression found by Zerilli\footnote{Up to a redefinition $a^{\ell m}\to a^{\ell m}/(\sqrt{4\pi})$, which is due to a different convention in (\ref{classicalST}).} (equation 20 in \cite{Zerilli1974}). Furthermore, for $Q=0$ we recover the ordinary Regge-Wheeler equation, as expected from the results of the previous section.

The last independent equation  $E_{t\theta}=0$ is satisfied identically with $h^{\ell m}_0$ and $h^{\ell m}_1$ (or $Q_{\rm axial}$) satisfying the above equations. The only last step is to solve (\ref{DHE1}) for $Q^{\ell m}_{\rm axial}$. Equation (\ref{DHE1}) can be recast as an ordinary Regge-Wheeler equation with the  transformations 
\bea
Q^{\ell m}_{\rm axial}&=&\left[1-\frac{4 k_1 Q^{2}\left(89 Q^{2}+5 r(-40 M+27 r)\right)}{5 r^{4}\left(Q^{2}+r(-2 M+r)\right)}\right]Z^{\ell m}_G,  \label{T1DH}\\
 a^{\ell m}&=&\left[1+\frac{ k_1(80Q^2-8Mr) }{r^4}\right]Z^{\ell m}_e \label{T2DH}\, ,
\eea
which yields
\bea
\left[ \tilde N^2  \frac{d^2}{dr^2}+\tilde N \tilde N'   \frac{d}{dr}+ (\omega^2- V^g_{\ell}(r) ) \right]Z^{\ell m}_G=\tilde C_4(\omega, r)Z^{\ell m}_e + C_5(\omega, r)\frac{dZ^{\ell m}_e}{dr}  \label{DHE2}\, ,
\eea
where $\tilde N = \sqrt{-\sigma N^2}$  is the usual function of the background metric  (\ref{metric0})-(\ref{eq:RNMetricCorrections}), and
\bea
V^g_{\ell}&=&\frac{\left(Q^{2}+r(-2 M+r)\right)\left(4 Q^{2}+r(-6 M+L r)\right)}{r^{6}} \\
&& -\frac{8 k_1 Q^{2}\left(426 Q^{4}+Q^{2} r(-1521 M+(394+83 L) r)+5 r^{2}\left(272 M^{2}-(139+31 L) M r+12\left(1+L\right) r^{2}\right)\right)}{5 r^{10}}\, ,\nonumber\\
\tilde C_4 & = & \frac{8  { i } \sqrt{\pi} {Q \omega}\left({Q}^{2}+{r}(-2 {M}+{r})\right)}{{r}^{5}}+\frac{32  { i }  { k_1 } \sqrt{\pi} {Q \omega}\left(529 Q^{4}+10(79 M-39 r)(2M-r) r^{2}+5 Q^{2} r(-374M+193 r)\right)}{5 {r}^{9}}\, ,
\eea
valid to first order in $k_1$.

\subsection{Electromagnetic perturbations}

By plugging (\ref{DHa}) and (\ref{DHh}) in the explicit expression that one gets for 
the perturbed semiclassical Maxwell equation (\ref{perturbM}), we obtain only one independent, nontrivial equation: $M^{\theta}=0$. By evaluating this equation in the gravitational and electromagnetic backgrounds  (\ref{DHg})-(\ref{DHA}), and taking into account the values of $h^{\ell m}_0(r)$, $h^{\ell m}_1(r)$ obtained in the previous section for the metric perturbations, we are able to find the following second-order ODE for the electromagnetic perturbation $a^{\ell m}$:
\bea
\left[D_1(r) \frac{d^2}{dr^2}+ D_2(r) \frac{d}{dr}+ (\omega^2- D_3(\omega,\ell,r) ) \right]a^{\ell m} =D_4(\omega, r)Q_{\rm axial}^{\ell m} + D_5(\omega, r)\frac{dQ_{\rm axial}^{\ell m}}{dr}  \label{DHM1}\, ,
\eea
where 
\bea
D_1&=& \frac{\left(Q^{2}+r(-2 M+r)\right)^{2}}{r^{4}}-\frac{8 k_1\left(Q^{2}+r(-2 M+r)\right)\left(48 Q^{4}+10 M(2 M-r) r^{2}+5 Q^{2} r(-19 M+5 r)\right)}{5 r^{8}} \nonumber \, ,\\
D_2 &=& \frac{2\left(-Q^{2}+M r\right)\left(Q^{2}+r(-2 M+r)\right)}{r^{5}}+\frac{8 k_1\left(Q^{2}+r(-2 M+r)\right)\left(398 Q^{4}+10 M(8 M-3 r) r^{2}+5 Q^{2} r(-139 M+42 r)\right)}{5 r^{9}} \nonumber\, ,\\
D_3&=&\frac{\left(4 Q^2+L r^2\right)\left(Q^2+r(-2 M+r)\right)}{r^6} \nonumber\\
&-& \frac{8k_1 \left[440 Q^8-10 Q^6 r(200 M+3(-27+L) r)+Q^4 r^2\left(2720 M^2+140(-14+L) M r-60(-5+L) r^2-97 r^4 \omega^2\right) \right]}{5r^{10}(Q^2+r(-2M+r))}\nonumber\\
&-& 8k_1\frac{ -5 Q^2 r^3\left(192 M^3+8(-17+5 L) M^2 r+2(7+3 L) r^3-8 M(r+2 l l r)^2+r^4(-43 M+24 r) \omega^2\right)}{5r^{10}(Q^2+r(-2M+r))}\nonumber\\
&-&\frac{16k_1  M(-2 M+r)\left(r^3 \omega^2+2 L(-2 M+r)\right)}{r^{5}(Q^2+r(-2M+r))}\, ,\nonumber\\
D_4 &=&-	\frac{i\left(-2+L\right) Q\left(Q^{2}+r(-2 M+r)\right)}{2 \sqrt{\pi} r^{5} w}+ \frac{2 i k_1\left(-2+L\right) Q\left(61 Q^{4}+60 M(2 M-r) r^{2}+5 Q^{2} r(-32 M+3 r)\right)}{5 \sqrt{\pi} r^{9} w}\, ,	  \nonumber\\
D_5 &=& 	-\frac{12  i k_1 \left(-2+L\right) Q\left(Q^{2}+r(-2 M+r)\right)^{2}}{\sqrt{\pi} r^{8} w}		\, ,	
\eea
and $L=\ell(\ell+1)$. In the classical limit $k_1=0$ our result recovers the expression found by Zerilli\footnote{Up to a redefinition $a^{\ell m}\to a^{\ell m}/(\sqrt{4\pi})$, which is due to a different convention in (\ref{classicalST}).} (equation 21 in \cite{Zerilli1974}).  To arrive at this expression we have made use of the classical result to get rid of higher-order derivative terms in the source: $k_1 \frac{d^2Q_{\rm axial}^{\ell m}}{dr^2}$, $k_1 \frac{d^3Q_{\rm axial}^{\ell m}}{dr^3}$, $k_1 \frac{d^4Q_{\rm axial}^{\ell m}}{dr^4}$, which is allowed to first order in $k_1$. 
 
Equation (\ref{DHM1}) can be recast as an ordinary Regge-Wheeler equation with the  transformations (\ref{T1DH})-(\ref{T2DH}), which yields
\bea
\left[ \tilde N^2  \frac{d^2}{dr^2}+\tilde N \tilde N'   \frac{d}{dr}+ (\omega^2- V^{e}_{\ell}(r) ) \right]Z^{\ell m}_e=\tilde D_4(\omega, r)Z^{\ell m}_G + D_5(\omega, r)\frac{dZ^{\ell m}_G}{dr}  \label{DHM2}\, ,	
\eea
where $\tilde N = \sqrt{-\sigma N^2}$ is the usual function of the background metric (\ref{metric0})-(\ref{eq:RNMetricCorrections}). On the other hand,
\bea
V^e_{\ell} &=& \frac{\left(4 Q^{2}+L r^{2}\right)\left(Q^{2}+r(-2 M+r)\right)}{r^{6}} \nonumber \\
&&+ \frac{8 k_1\left(-1452 Q^{6}+30 M(2 M-r) r^{3}\left(5 M+\left(-2+L\right) r\right)+Q^{4} r(5550 M+(-2290+127 L) r)\right)}{5r^{10}}\nonumber\\ &&	+ 8 k_1\frac{Q^{2} r^{2}\left(-1106 M^{2}+(922-59 L) M r+6(-31+5 L) r^{2}\right)}{r^{10}}\, , \\
\tilde D_4 &=& -\frac{i\left(-2+L\right) Q\left(Q^{2}+r(-2 M+r)\right)}{2 \sqrt{\pi} r^{5} \omega} +k_1 \frac{4 {i}\left(-2+L\right) Q\left(28 Q^{4}+25 {M}(2 {M}-r) r^{2}+5 Q^{2} r(-14 {M}+r)\right)}{5 \sqrt{\pi} r^{9} \omega} \, ,	
\eea
Notice that for $Q=0$ we recover our previous result, Eq. (\ref{eq:Regge-Wheeler}) with effective potential (\ref{eq:AxialPotentialQuantum}). Eq. (\ref{DHM2}) thus generalizes Eq. (\ref{eq:Regge-Wheeler}) when $Q\neq 0$.

\subsection{Master equations} \label{masterdh}

If we define the tortoise radial coordinate by $d r=\tilde N(r) dr_*$, then electromagnetic and gravitational linear perturbations on the background spacetime (\ref{DHg})-(\ref{DHA}) of Sec. \ref{DHsolution} propagate according to a pair of coupled  wave equations:
\bea 
\left[  \frac{d^2}{dr_*^2}+ (\omega^2- V^{g}_{\ell}(r) ) \right]Z^{\ell m}_G=\tilde C_4(\omega, r)Z^{\ell m}_e + \frac{C_5(\omega, r)}{\tilde N(r)}\frac{dZ^{\ell m}_e}{dr_*}\, ,	\label{MDH1}\\
\left[  \frac{d^2}{dr_*^2}+ (\omega^2- V^{e}_{\ell}(r) ) \right]Z^{\ell m}_e=\tilde D_4(\omega, r)Z^{\ell m}_G + \frac{D_5(\omega, r)}{\tilde N(r)}\frac{dZ^{\ell m}_G}{dr_*} \, . \label{MDH2}
   \eea 
All quantum corrections are encoded in the effective potential as well as in the source terms.  

It is not difficult to check that, in a neighborhood around the BH horizon (\ref{horizondh}) and at spatial infinity $r\to \infty$,   the effective potential and source terms vanish to first order in $k_1\sim \hbar$. Consequently, the two linearly independent solutions   of each of the equations above can be still taken such that:
\bea
   Z^{\ell m}_{\lambda} &\underset{r\to r_H}{\sim} & A_{\lambda\omega\ell} \, e^{-i\omega r_{*H}} + B_{\lambda\omega\ell} \, e^{i\omega r_{*H}}\, ,\\
    Z^{\ell m}_{\lambda} &\underset{r\to \infty}{\sim} &  C_{\lambda\omega\ell} \, e^{-i\omega r} + D_{\lambda\omega\ell} \, e^{i\omega r}\, ,
\eea
where $\lambda\in\{G, e\}$ and, to first order in $k_1$, the tortoise coordinate reads
   \bea
   r_*(r)= \frac{r r_--r r_++r_-^{2} \log [r-r_-]-r_+^{2} \log [r-r_+]}{r_--r_+} \hspace{12cm} \nonumber\\
   +\frac{k_1}{5 } \left[ \frac{196}{r}-\frac{152(r_-+r_+) \log [r]}{r_- r_+}+ \left[ \frac{64}{r_-}+\frac{44r_-}{r_+^2}-\frac{8}{r_+}+\frac{220}{r_+-r_-} \right]\log (r-r_-)+ \left[ \frac{64}{r_+}+\frac{44r_+}{r_-^2}-\frac{8}{r_-}+\frac{220}{r_--r_+} \right]\log (r-r_+) \right]\, . \nonumber
   \eea

\section{A charged black hole with Euler-Heisenberg corrections}
\label{sec:EH}

In this and in the next sections we want to compare the charged BH solution that we have  obtained  in Sec. \ref{DHsolution} from the Drummond-Hathrell semiclassical corrections, Eqs.  (\ref{DHg})-(\ref{DHA}), with the solution that one may get by working instead with the Euler-Heisenberg corrections in (\ref{gammaaction}).

Again, we evaluate the functional derivatives (\ref{variations}) using \texttt{xAct} in order to obtain explicit expressions for $\langle T_{ab}^{(1)} \rangle$, $\langle J_a^{(1)} \rangle$,  generated by the corrections  $A$, $B$ of (\ref{gammaaction}). The output is terribly cumbersome and not particularly interesting. Then, we look  for static and spherically symmetric solutions  of (\ref{fieldeqns}) by working with the metric Ansatz 
\be 
ds^2=N(r)\sigma(r) dt^2+\frac{dr^2}{N(r)}+r^2d\theta^2+r^2\sin^2\theta d\phi^2, \label{Nsigma}
\ee
and the electromagnetic potential 
\be 
A(r)=\big(A_0(r),0,0,0\big).
\ee 
Using these expressions in the semiclassical field equations (\ref{fieldeqns}) leads to four independent, coupled differential equations, three from the gravitational sector (the $tt$, $rr$ and $\theta\theta$ components) and one from the electromagnetic one (the $t$ component). To solve this system of coupled differential equations we write
\begin{align}
\label{eq:Euler-Heisenbergansatz}
N(r)&=\bigg(1-\frac{2M}{r}+\frac{Q^2}{r^2}\bigg)\big(1+k_2 \delta N\big),\nonumber\\
\sigma(r)&=-1+k_2\delta \sigma,\nonumber\\
A_0(r)&=\frac{Q}{\sqrt{4\pi}r}+k_2\delta A,
\end{align}
where  $k_2$ was defined in equation \eqref{k1k2constants}. The $tt$ and $rr$ components of the Einstein equations, expanded to order $\mathcal{O}(k_2^2)$, give
    \begin{align}
& 4 \sqrt{\pi } r^6 Q  \delta A' - r^4\left[ r \delta N' (Q^2-2 M
   r+r^2)+\delta N (r^2-Q^2)+ Q^2 \delta\sigma \right]+\frac{3
   Q^4}{\pi}=0,\\
&r^4 \left[Q^2 \delta\sigma+ r \left(\left(-2 M r+Q^2+r^2\right) \left(\delta N'-\delta\sigma'\right)-4
   \sqrt{\pi} Q r \delta A'\right)+\delta N \left(r^2-Q^2\right)\right]-\frac{3 Q^4}{\pi}=0,
    \end{align}
and the nonzero component of the Maxwell sector yields 
\begin{equation}
\pi  r^5 \left[4 \sqrt{\pi } r \left(r \delta A''+2 \delta A'\right)-Q \delta\sigma'\right]-16 Q^3=0.
\end{equation}
This coupled system of three differential equations can be solved in full closed form.  If we demand that the solution approaches the Minkowski metric at spatial infinity (i.e. $\delta N(r)$, $\delta \sigma(r)$, $\delta A(r)\to 0$ as $r\to\infty$), and we fix the mass and charge of the resulting solution to be $M$ and $Q$, respectively, (by identifying the $r^{-1}$ and $r^{-2}$ prefactors in the asymptotic expansion of the lapse function)  one obtains
\beq
\label{eq:Euler-Heisenbergcorr}
\delta N(r)&=&\frac{Q^4}{5\pi r^4(Q^2+r(-2M+r))},\nonumber\\
\delta \sigma(r)&=&0,\nonumber\\
\delta A(r)&=&\frac{Q^3}{5\pi^{3/2}r^5}\,,
\eeq
which also satisfies the remaining fourth differential equation of the system (the $\theta\theta$ component of the semiclassical Einstein equations). 

These results agree with the ones obtained in equations (54) and (55) of Ref.~\cite{Ruffini:2013hia} with the identification $Q\to-Q/\sqrt{4\pi}$, and  $E_c\to m_e^2/e=m_e^2/\sqrt{4\pi\alpha}$. However, they differ by a factor of $2/\pi$ from the ones obtained in Ref.~\cite{Abbas:2023nra}.

Overall, the semiclassical metric and electromagnetic potential solutions are 
\bea
ds^2&=&-\left(1-\frac{2M}{r}+\frac{Q^2}{r^2}+k_2 \frac{Q^4}{5\pi r^6} \right) dt^2+\frac{dr^2}{1-\frac{2M}{r}+\frac{Q^2}{r^2}+k_2 \frac{Q^4}{5\pi r^6}}+r^2 (d\theta^2+\sin^2\theta d\phi^2)\,, \label{EHg}\\
A(r)&=&\frac{Q}{\sqrt{4\pi}r}\left[1+k_2\frac{2Q^2}{5\pi r^4} \right]dt\, . \label{EHA}
\eea
 To leading-order in the coupling constant $k_2$, the (highest) roots of $g_{tt}(r_H)=0$ and $g^{-1}_{rr}(r_H)=0$ lead to
\bea
r_H=r_+ + k_2 \frac{r_-^2}{5\pi r_+^2(r_- - r_+)}+\mathcal O(k_2^2)\, , \label{horizoneh}
\eea
where $r_{\pm}=M\pm \sqrt{M^2-Q^2}$ are the  horizons of the classical Reissner-Nordstr\"om BH. Therefore, this spacetime background contains a BH horizon at $r=r_H$.

\section{Axial perturbations on a charged black hole with Euler-Heisenberg corrections}

We will now derive the wave equations for the propagation of electromagnetic and gravitational linear perturbations on the spacetime background obtained in Sec. \ref{sec:EH}, given in (\ref{EHg})-(\ref{EHA}).

For $Q=0$ this background reduces to the ordinary, neutral Schwarzschild spacetime. Since the Euler-Heisenberg corrections only produce  quadratic  (or higher-order) polynomials  in $F_{ab}$ on the RHS of both equations in (\ref{fieldeqns}),  linear perturbations  around a background with  $F_{ab}=0$  do not produce any deviation with respect to the classical case. Therefore, unlike the Drummond-Hathrell case described in Sec. \ref{sec:partialproblem}, the problem is trivial if $Q=0$  (i.e. perturbations totally decouple and the problem reduces to that of solving classical Maxwell equations in a neutral background). For this reason, we will focus directly on the most general case, $Q\neq 0$.

To solve this problem we follow exactly the same steps as in Sec. \ref{sec:DH}. First of all, we derive the specific semiclassical field equations (\ref{fieldeqns}) that one obtains from the effective action (\ref{gammaaction}) by taking suitable field variations (\ref{variations}) of the $A$, $B$ corrections. This is carried out using \texttt{xAct}. We then linearize this answer by using the decomposition (\ref{decomp}) for both the metric and electromagnetic potential, using the Ansatze (\ref{DHh}) and (\ref{DHa}), respectively, and working again with \texttt{xAct}. This produces  lengthy tensorial equations, that we denote by (\ref{perturbE}) and (\ref{perturbM}). Our final task is to determine  the three unknown functions $h_{0,\ell m}(r)$, $h_{1,\ell m}(r)$, $a^{\ell m}$ in (\ref{DHh}) and (\ref{DHa}) by solving these equations.

\subsection{Gravitational perturbations} 

The problem is very similar to the one described in Sec \ref{sec:DH}. If we plug (\ref{DHa}) and (\ref{DHh}) in the explicit expressions that one gets for the perturbed semiclassical Einstein equation, (\ref{perturbE}), we get four independent, nontrivial equations: $E_{t\theta}=0$, $E_{r\theta}=0$, $E_{\theta\theta}=0$, $E_{\theta\phi}=0$. The equation $E_{\theta\phi}=0$ can be solved for $h^{\ell m}_0(r)$ in terms of $h^{\ell m}_1(r)$ and its first derivative.  Then $E_{\theta\theta}=0$ is satisfied as an identity, and we remain with two independent equations, $E_{t\theta}$ and $E_{r\theta}$, and one unknown function $h^{\ell m}_1(r)$. Doing the transformation $h^{\ell m}_1(r)=Q^{\ell m}_{\rm axial}(r)\frac{r}{1-\frac{2M}{r}+\frac{Q^2}{r^2}}$, the equation $E_{r\theta}=0$ leads to a second-order differential equation for $Q^{\ell m}_{\rm axial}(r)$:
\bea
\left[C_1(r) \frac{d^2}{dr^2}+ C_2(r) \frac{d}{dr}+ (\omega^2- C_3(\omega,\ell,r) ) \right]Q^{\ell m}_{\rm axial}=C_4(\omega, r)a^{\ell m}  \label{EHE1}\, ,	
\eea
where, up to $\mathcal O(k_2^2)$, we have
\bea
C_1 &=&\frac{\left(Q^{2}+r(-2 M+r)\right)^{2}}{r^{4}}+\frac{k_2 Q^{4}\left(Q^{2}+r(-2 M+r)\right)}{5 \pi r^{8}} \, ,	\nonumber\\
C_2 &=& -\frac{2\left(Q^{2}-M r\right)\left(Q^{2}+r(-2 M+r)\right)}{r^{5}}-\frac{2 k_2 Q^{4}\left(7 Q^{2}+r(-16 M+9 r)\right)}{5 \pi r^{9}}\nonumber\, ,	\\
C_3& =& \frac{\left(Q^{2}+r(-2 M+r)\right)\left(4 Q^{2}+r(-6 M+L r)\right)}{r^{6}}-  \frac{k_2 Q^{4}\left(20 Q^{4}+2 Q^{2} r(-46 M+25 r)+r^{2}\left(108 M^{2}-120 M r+34 r^{2}-\omega^{2} r^{4}\right)\right)}{5 \pi r^{10}\left(Q^{2}+r(-2 M+r)\right)} \nonumber\, ,	\\
C_4 &=&\frac{8 {i} \sqrt{\pi} Q \omega \left(Q^{2}+r(-2 M+r)\right)}{r^{5}}- \frac{8 {i} k_2 Q^{5} \omega }{5 \sqrt{\pi} r^{9}}\nonumber \, ,	
\eea
and $L=\ell(\ell+1)$. For $k_2=0$ we recover the classical limit of Zerilli\footnote{Up to a redefinition $a^{\ell m}\to a^{\ell m}/(\sqrt{4\pi})$, which is due to a different convention in (\ref{classicalST}).} (see equation 20 in \cite{Zerilli1974}). In addition, for $Q=0$ we recover the well-known Regge-Wheeler equation  for gravitational  perturbations on a Schwarzschild background.

We now rewrite the above equation as a Regge-Wheeler equation
\bea
\left[ \tilde N^2  \frac{d^2}{dr^2}+\tilde N \tilde N'   \frac{d}{dr}+ (\omega^2- V^g_{\ell}(r) ) \right]Z^{\ell m} _G=\tilde C_4(\omega, r)Z_e^{\ell m}   \label{EHE2}\, ,	
\eea 
for some $Z_G$ and $Z_e$, and where $\tilde N = \sqrt{-\sigma N^2}$ is the usual function of the background metric (\ref{Nsigma})-(\ref{eq:Euler-Heisenbergcorr}).  This is accomplished with the transformation 
\bea
Q_{\rm axial}&=&\left[1-\frac{k_2 Q^{4}}{5 \pi r^{4}\left(Q^{2}+r(-2 M+r)\right)}\right]Z_G\, , \label{T1EH} \\
a^{\ell m}&=&\left[1+\frac{ k_2 Q^2 }{\pi r^4}\right]Z^{\ell m}_e, \label{T2EH}
\eea 
which leads to
\bea
 V^g_{\ell}&=& \frac{\left(Q^{2}+r(-2 M+r)\right)\left(4 Q^{2}+r(-6 M+L r)\right)}{r^{6}}-\frac{k_2 Q^{4}\left(-12 Q^{2}+r\left(22 M-\left(8+L\right) r\right)\right)}{5 \pi r^{10}} \, ,\label{VGEH}	\\
\tilde C_4(\omega, r)& =& \frac{8 {i} \sqrt{\pi} Q\left(Q^{2}+r(-2 M+r)\right) \omega }{r^{5}}+\frac{8 \dot{i} k_2\left(6 Q^{5}+5 Q^{3} r(-2 M+r)\right) \omega }{5 \sqrt{\pi} r^{9}} \, . \label{C4EH}	
\eea

\subsection{Electromagnetic perturbations} 

We repeat the same steps of previous sections. Namely, we plug (\ref{DHa}) and (\ref{DHh}) in the specific result that we obtain for the perturbed semiclassical Maxwell equation (\ref{perturbM}) using \texttt{xAct}, and we obtain only one independent, nontrivial equation: $M^{\theta}=0$. We evaluate this equation in the spacetime and electromagnetic backgrounds of Sec. \ref{sec:EH}, given in (\ref{EHg})-(\ref{EHA}). Taking into account the values of $h^{\ell m}_0(r)$, $h^{\ell m}_1(r)$ obtained in the previous subsection for the metric perturbations, we are able to find the following second-order ODE for the electromagnetic perturbation $a^{\ell m}$:
\bea
\left[D_1(r) \frac{d^2}{dr^2}+ D_2(r) \frac{d}{dr}+ (\omega^2- D_3(\omega,\ell,r) ) \right]a^{\ell m} =D_4(\omega, r)Q_{\rm axial}^{\ell m}  \label{DHM1} \, ,	
\eea
where  now
\bea
D_1 &=& \frac{\left(Q^{2}+r(-2 M+r)\right)^{2}}{r^{4}} -\frac{M { k_2 }\left(Q^{2}-2 M r+r^{2}\right)\left(9 Q^{4}-20 M Q^{2} r+10 Q^{2} r^{2}\right)}{5 \pi r^{8}} \, ,	\\
D_2 &=& =\frac{2\left(-Q^{2}+M r\right)\left(Q^{2}+r(-2 M+r)\right)}{r^{5}}+\frac{2 k_2 Q^{2}\left(27 Q^{2}-50 M r+20 r^{2}\right)\left(Q^{2}+r(-2 M+r)\right)}{5 \pi r^{9}} \, ,	\\
D_3 &=& \frac{20 \pi Q^{2} r^{2}\left(Q^{2}+r(-2 M+r)\right)^{2}+5 L \pi r^{4}\left(Q^{2}+r(-2 M+r)\right)^{2}}{5 \pi r^{8}\left(Q^{2}+r(-2 M+r)\right)}\\
& &+ \frac{M { k_2 }\left(25 L Q^{2}\left(Q^{2}+r(-2 M+r)\right)^{2}+Q^{2} r^{4}\left(11 Q^{2}+10 r(-2 M+r)\right) \omega^{2}\right)}{5 \pi r^{8}\left(Q^{2}+r(-2 M+r)\right)}\, ,	\\
D_4 &=&  \frac{{i}\left(-2+L \right)  { Q  }\left(Q^{2}+r(-2 M+r)\right)}{2 \sqrt{\pi} \omega r^{5}}+\frac{{i}\left(-2+L \right) Q^{5}   { k_2 }}{10 \pi^{3 / 2} \omega r^{9}}\, ,	
\eea
and $L=\ell(\ell+1)$. Again, for $k_2=0$ we recover the classical limit of Zerilli, up to a redefinition of $a^{\ell m}$ (see equation (21) in \cite{Zerilli1974} and our footnote 7). Furthermore, for $Q=0$ we recover the well-known Regge-Wheeler equation  for electromagnetic  waves on a Schwarzschild background.

Performing the change of variables (\ref{T1EH}), (\ref{T2EH}), we can cast the above result in terms of a Regge-Wheeler wave equation:
\bea
\left[ \tilde N^2  \frac{d^2}{dr^2}+\tilde N \tilde N'   \frac{d}{dr}+ (\omega^2- V^{e}_{\ell}(r) ) \right]Z^{\ell m}_e=\tilde D_4(\omega, r)Z_{G}^{\ell m}   \label{EHM2}\, ,	
\eea
where
\bea
V_{\ell}^e & =&\frac{\left(4 Q^{2}+L r^{2}\right)\left(Q^{2}+r(-2 M+r)\right)}{r^{6}}- k_2\frac{ Q^{2}\left(96 Q^{4}+5(2 M-r) r^{2}(48 M+(-20+7 L) r)-4 Q^{2} r(110 M+(-50+9 L) r)\right)}{5 \pi r^{10}} \, ,\label{VeEH}	\\
\tilde D_4 & =& -\frac{{i}\left(-2+L\right) Q\left(Q^{2}+r(-2 M+r)\right)}{2 \sqrt{\pi} \omega r^{5}}-k_2\frac{{i}\left(-2+L\right) Q^{3}\left(6 Q^{2}+5 r(-2 M+r)\right) }{10 \pi^{3 / 2} \omega r^{9}}  \label{D4EH}\, .
\eea

\subsection{Master equations}\label{mastereh}

To conclude, we can write the linearized Einstein and Maxwell coupled system of  equations, with one-loop Euler-Heisenberg corrections, as a pair of two coupled Regge-Wheeler equations for the variables $Z_{G}^{\ell m}$ and $Z_{e}^{\ell m}$:
\bea 
\left[  \frac{d^2}{dr_*^2}+ (\omega^2- V^{g}_{\ell}(r) ) \right]Z^{\ell m}_G=\tilde C_4(\omega, r)Z^{\ell m}_e \, ,	\label{MEH1} \\
\left[  \frac{d^2}{dr_*^2}+ (\omega^2- V^{e}_{\ell}(r) ) \right]Z^{\ell m}_e=\tilde D_4(\omega, r)Z^{\ell m}_G \, . \label{MEH2}\
   \eea 
   where the tortoise coordinate is given by $d r=\tilde N(r) dr_*$. To first order in $k_2$, the integration yields
   \bea
   r_*(r)&=& \frac{r r_--r r_++r_-^{2} \log [r-r_-]-r_+^{2} \log [r-r_+]}{r_--r_+}\\
   &&-\frac{k_2}{5 \pi} \left[ -\frac{1}{r}+\frac{-\frac{r_+^{2}}{r-r_-}+\frac{r_-^{2}}{-r+r_+}}{(r_--r_+)^{2}}+\frac{2(r_-+r_+) \log [r]}{r_- r_+}+\frac{2 r_+^{3}(-2 r_-+r_+) \log [r-r_-]-2 r_-^{3}(r_--2 r_+) \log [r-r_+]}{r_-(r_--r_+)^{3} r_+} \right]\, , \nonumber
   \eea
   where $r_{\pm}=M\pm \sqrt{M^2-Q^2}$ are the two classical horizons for a Reissner-Nordstr\"om BH. Similar to what happens in the Drummond-Hathrell case,  in a neighborhood around the BH horizon (\ref{horizoneh}) and at spatial infinity $r\to \infty$,   the effective potential and source terms vanish to first order in $k_2\sim \hbar$.

\section{Conclusions and final remarks}
\label{sec:Conclusions}

The advent of gravitational-wave astronomy  has put BHs in the spotlight.  In particular, gravitational interferometers can extract the ringdown signal of solar-mass binary BH mergers, from which we can study the  physics underlying BH dynamics via the analysis of QNM frequencies. According to classical general relativity, for BHs these frequencies can only depend on three parameters: mass, spin and electric charge.  Current expectations for high precession measurements in future generation gravitational-wave interferometers motivates us to take a step forward and to calculate quantum corrections for BHs and their imprints in QNM frequencies. 

Even though the theory of quantum fields in curved spacetime is a mature field of research that dates back to as early as the 1960s, the intrinsic difficulties inherent in this framework  complicates  making significant progress in practical calculations. In particular, the space of solutions of the semiclassical Einstein equations is still pretty much unexplored. Furthermore, linear perturbation theory has not even been addressed in this framework to the best of our knowledge.  

Much of the difficulties in the search for solutions to the semiclassical field equations is owed to the problem of renormalization in curved spacetime. In practical applications, we do not have a systematic way to renormalize the vacuum expectation value of the stress-energy tensor of quantum fields, and to solve these equations for a sufficiently wide family of  spacetime metrics, not even using numerical methods. In this work we have opted to work with perturbative expansions of the one-loop effective action that can be obtained using heat kernel techniques. These approximations miss the physical details of the quantum state, but they still provide leading-order quantum corrections to the classical action of general relativity, which are  expected to dominate for weak background fields. 

In the present work we used the Drummond-Hathrell \cite{Drummond:1979pp} and Euler-Heisenberg \cite{DUNNE_2005} approximate expressions for the one-loop effective action to derive  static, spherically symmetric solutions of the full  Einstein-Maxwell semiclassical equations. We have been able to find solutions that  are exact to leading-order in  Planck's constant. Our  results can be found in (\ref{DHg})-(\ref{DHA}) and  (\ref{EHg})-(\ref{EHA}), respectively. The latter case agrees with previous studies \cite{Ruffini:2013hia}. Interestingly, the quantum corrections  do not add more ``hair'' to the classical BHs (at least, in spherical symmetry). More precisely, after imposing asymptotic flatness and the values  for the mass  $M$ and electric charge  $Q$ of the resulting compact object, all free constants of integration vanish.  The corrections are entirely determined by the two parameters $M$ and $Q$. 

According to general relativity, BHs react when they are subject to small perturbations, as a consequence of which they emit gravitational  radiation with a characteristic frequency spectrum. If they are electrically charged, they can also emit electromagnetic radiation with an associated spectrum. We have studied here the propagation of both electromagnetic and gravitational linear perturbations around the  semiclassical solutions (\ref{DHg})-(\ref{DHA}) and  (\ref{EHg})-(\ref{EHA}). We have first addressed the $Q=0$ case for (\ref{DHg})-(\ref{DHA}), since in this case the  Maxwell sector still receives quantum corrections (the gravitational sector does not). In particular,  we have derived the equation describing the propagation of both axial and polar electromagnetic perturbations subject to vacuum polarization effects, which can be consulted in (\ref{eq:AxialPotentialQuantum}) and (\ref{eq:PolarPotentialQuantum}) respectively. To complete the analysis, we also computed the spectrum of characteristic frequencies for the fundamental tone and the first few angular momentum values, written in equations (\ref{axial1})-(\ref{polar2}). We verified that quantum corrections scale linearly with Planck's constant, and that the classical isospectrality is lost. 

After this, we addressed the full perturbation problem for $Q\neq 0$. For axial perturbations, and for both backgrounds (\ref{DHg})-(\ref{DHA}) and  (\ref{EHg})-(\ref{EHA}), we have obtained   the relevant coupled pair of Regge-Wheeler  master equations that govern the evolution of these linear waves. Main results can be seen in (\ref{MDH1})-(\ref{MDH2}) for Drummond-Hathrell corrections and (\ref{MEH1})-(\ref{MEH2}) for Euler-Heisenberg corrections. In the classical limit we recover well-known results \cite{Zerilli1974,Chandrasekhar:1985kt}. Our findings extend these references when  vacuum polarization effects of electron-positron pairs are taken into account. Polar perturbations, on the other hand, are technically much more involved due to the existence of many more field variables and issues with gauge invariance. We found significant difficulties in solving this problem, particularly because our {\it Mathematica} notebooks were unable to produce results, even with the use of the  powerful packages of \texttt{xAct}. This study is left  for a future work.

In the  future we plan to calculate the dominant QNM frequencies for the system of equations (\ref{MDH1})-(\ref{MDH2}) and (\ref{MEH1})-(\ref{MEH2}), for the Drummond-Hathrell  and Euler-Heisenberg semiclassical solutions, respectively. At first sight, one may think this is a straightforward exercise by using standard numerical techniques. However,  {the presence of $\hbar$ makes the problem considerably more complicated}, a difficulty that gets particularly enhanced when $Q\neq 0$. As  discussed in Sec. \ref{qnmsection}, since the actual value of $\hbar$ is dozens of orders of magnitude below machine precision,  one is forced to do the numerical calculation using several, higher artificial values of $\hbar$, and then one  expects to extract the actual numerical result for the QNM spectra by  linear extrapolation. However, we cannot use arbitrarily big values of  $\hbar$ either, otherwise perturbation theory breaks down and, among other complications,  the corrections need not follow a linear dependence. On the other hand, if the chosen values of $\hbar$ are still too small, one may not have sufficient numerical precision to infer the impact of $\hbar$ on the classical spectra, and quantum corrections remain buried in numerical errors.

This issue   arose already in the $Q=0$ case, but it was still tractable  (see footnote 7). {Because a nonzero electric charge $Q\neq 0$ couples both equations in   (\ref{MDH1})-(\ref{MDH2}) and (\ref{MEH1})-(\ref{MEH2}),  the problem gets  qualitatively and significantly more complicated than the $Q=0$ case}. For relatively big values of $\hbar$ (but still below $M^2$) we found that, using the direct integration method (Appendix A),    quantum corrections seem to affect  the frequency spectra quadratically, i.e. to order $O(\hbar^2)$. Taken at face value, this would mean that, to leading-order in the perturbative (quantum) framework, the QNM frequency spectra remains unaffected. However, this is not apparent from equations   (\ref{MDH1})-(\ref{MDH2}) and (\ref{MEH1})-(\ref{MEH2}), which include explicit $O(\hbar^1)$ corrections. On the other hand, when we decreased the size of $\hbar$ in order to approach the linear regime, we  found  convergence problems   against variations of the parameters used. The determination of the QNM frequency spectra when  $Q\neq 0$ is technically more demanding  and deserves a separate study.

As remarked in the Introduction, effective field theories have a broad range of validity, which make them useful to test fundamental physics. The most general low-energy limit of quantum gravity and electrodynamics is expected to have the form of (\ref{gammaaction}) for some unknown coefficients $a$, $b$, $c$ etc. From a theoretical viewpoint, their specific value depends on the details of the particular UV completion of the theory. Although calculations would become more tedious, our results could be easily extended for free parameters in the action (\ref{gammaaction}), so that their theoretical prediction could, in principle, be tested with gravitational-wave observations.\\

{\bf \em Acknowledgments.} 
We thank Vitor Cardoso for several useful comments and insights during the course of this work, as well as for a careful reading of the article. ADR thanks Jose Navarro-Salas for  comments and feedback. EE wishes to thank Maarten van de Meent for helpful comments and discussions. EE would also like to thank Rodrigo Panosso Macedo and David Pere\~niguez for useful discussions while preparing this manuscript. 
ADR is supported through a M. Zambrano grant (ZA21-048) with reference UP2021-044 from the Spanish Ministerio de Universidades, funded within the European Union-Next Generation EU. This work is also supported by the Spanish Grant PID2020-116567GB-C21 funded by MCIN/AEI/10.13039/501100011033.
EE acknowledges support from the Villum Investigator program supported by the VILLUM Foundation (grant no. VIL37766) and the DNRF Chair program (grant no. DNRF162) by the Danish National Research Foundation. This project has also received funding from the European Union's Horizon 2020 research and innovation program under the Marie Sklodowska-Curie grant agreement No 101131233.

\begin{appendix}

\section{Numerical method}

In this appendix, we briefly review  the  direct integration method for computing the QNM spectra,  which we used here to derive the numerical values  in Tables \ref{table:FundamentalModeAxial} and \ref{table:FundamentalModePolar}. These tables contain the electromagnetic quasinormal (QN) frequencies for $Q=0$ when Drummond-Hathrell corrections are considered, corresponding to equations (\ref{eq:Regge-WheelerA}) and (\ref{eq:Regge-WheelerP}) for axial and polar perturbations, respectively (recall that the classical gravitational QN spectra do not get affected when $Q=0$). This method can also be  applied to calculate the QNM spectra of equations \eqref{MDH1} - \eqref{MDH2} (Drummond-Hathrell corrections) and  \eqref{MEH1} - \eqref{MEH2} (Euler-Heisenberg corrections), valid when $Q\neq 0$. For completeness, we will treat here the general description for any $Q$. On the other hand, for details on the pseudospectrum technique, which we only used in Sec. \ref{qnmsection},  we refer to the original papers  \cite{Macedo_2018}, \cite{Macedo_2020}, which include all the necessary mathematical machinery.

 As mentioned  in Secs. \ref{masterdh} and \ref{mastereh}, in a neighborhood of the BH horizon and at spatial infinity,   the effective potentials and source terms vanish to first order in $\hbar$. Therefore, the boundary conditions that define the QNMs are still formally equal as in the classical theory: 
 \bea Z^{\ell m}_{G,e} &\underset{r\to r_H}{\sim} &  e^{-i\omega r_{*H}}\, ,\\ 
 Z^{\ell m}_{G,e} &\underset{r\to \infty}{\sim} &  e^{i\omega r_{*H}}\,, 
 \eea
but with $r_{*H}\equiv r_*(r_H)$  the (quantum-corrected) tortoise coordinate evaluated at the horizon radius. To reach sufficient numerical precision in the calculation, we need to include higher-order terms in the above asymptotic relations. We can do this by using Frobenius expansions of the form\footnote{We have  omitted the $m$ label in the coefficients since there is no actual dependence on the azimuthal number due to spherical symmetry of the background.}
\bea Z^{\ell m}_{G,e} &=&e^{-i\omega r_{*H}} \sum_{i=0}^{ordH} f^{G,e}_{\ell\, i} (r-r_H)^i\,, \label{bounh}\\ Z^{\ell m}_{G,e} &=&e^{i\omega r_{*H}} \sum_{i=0}^{ordI}  \frac{g^{G,e}_{\ell\, i}}{r^i}\, . \label{bouni} \eea
 The coefficients $f^{G,e}_{\ell\, i}$ and $g^{G,e}_{\ell\, i}$ are obtained by solving iteratively and  order by order the field equations  \eqref{MDH1} - \eqref{MDH2} (Drummond-Hathrell corrections) or  \eqref{MEH1} - \eqref{MEH2} (Euler-Heisenberg corrections), together with the initial value $g^{G,e}_{\ell\, 0}=f^{G,e}_{\ell\, 0}=1$. Once all  these coefficients are calculated, we  integrate twice the field equations,   from the horizon and from infinity, using (\ref{bounh}) and (\ref{bouni}) and their derivatives as  boundary conditions.   We then impose a $C^1$ matching of the two numerical results  at some intermediate but otherwise arbitrary value of the radius, $r_{int}$. This matching condition severely constrains  the possible values that $\omega$ can take, and produces the QNM frequency spectra.

For $Q=0$ and Drummond-Hathrell corrections, the results found for the allowed values of $\omega$ are shown in Tables \ref{table:FundamentalModeAxial} and \ref{table:FundamentalModePolar} for $\ell =1, 2$ and several values of the coupling constant $\xi^2\sim k_1$ (see footnote 7 for details). We used $ordH=11$ and $ordI=9$ for the truncation of the Frobenius expansions (\ref{bounh}) and (\ref{bouni}).  
 To avoid the coordinate singularity at the horizon radius, we need to  integrate slightly away from this location. When $Q=0$ the BH horizon (\ref{horizondh}) reduces to the ordinary Schwarzschild radius, $r_H=2M$, so we locate the effective horizon at $r_H^{\epsilon}:=2M(1+\epsilon)$. Similarly, we place spatial infinity at some finite value  $r_{\infty}<\infty$. We checked the convergence of our calculations against variations of $\epsilon$,  $r_{\infty}$, and $r_{int}$. In particular, the  digits shown in Tables \ref{table:FundamentalModeAxial} and \ref{table:FundamentalModePolar} are robust for $\epsilon\in [10^{-2}, 10^{-4}]$, $r_{\infty}\in [20, 40]$, and $r_{int}\in [r_H^{\epsilon}, r_{\infty}]$.

\end{appendix}

\bibliography{references}

\end{document}